\documentclass[%
 reprint, 
superscriptaddress,
 amsmath,amssymb,
 aps, physrev,
]{revtex4-2}

\usepackage{graphicx}% Include figure files
\usepackage{dcolumn}% Align table columns on decimal point
\usepackage{bm}% bold math
\usepackage{braket}
\usepackage{amsfonts}
\usepackage{url}
\usepackage{esint}
\usepackage{hyperref}
\begin{document}

\preprint{APS/123-QED}

\title{\textbf{Analytical topological invariants for 2D non-Hermitian phases using Morse theory} 
}% 

\author{Cameron Gibson}
\email{camerongibson@tamu.edu}
\affiliation{Department of Physics and Astronomy, Rice University, Houston, Texas 77005, USA}
\affiliation{Center for Theoretical Biological Physics, Rice University, Houston, Texas 77005, USA}
\affiliation{George P. \& Cynthia Woods Mitchell Institute for Fundamental Physics and Astronomy,
Texas A\&M University, College Station, TX 77843, USA}
\author{Evelyn Tang}%
\email{e.tang@rice.edu}
\affiliation{Department of Physics and Astronomy, Rice University, Houston, Texas 77005, USA}
\affiliation{Center for Theoretical Biological Physics, Rice University, Houston, Texas 77005, USA}

\date{\today}% It is always \today, today,

\begin{abstract}
As energy dissipation and gain are ubiquitous in the real world, such phenomena demand the generalization of Hermitian methods such as the analysis of topological properties for non-Hermitian systems. However, as non-Hermitian systems typically contain more degrees of freedom, this poses a challenge for analytical approaches to understand their topology and invariants. In this work, we analytically calculate the 2D Zak phase for a 2D non-Hermitian SSH-type Hamiltonian that supports a rich structure and edge currents. Closed-form expressions for eigenstates and divisions of the phase diagram are obtained, including for regions in the phase diagram where different types of exceptional points exist. We use Morse theory to determine the topology of exceptional points in momentum space. Although the band structure breaks down at exceptional points, we show that a specific phase-based topological invariant remains well-defined. Furthermore, our work yields an analytic derivation for counting edge states in the Hermitian limit. These results provide new conceptual and analytical tools for the study of complex topological systems.
\end{abstract}

\maketitle

%\tableofcontents

\section{Introduction}
Topology has been very successful in predicting new phases of matter across different platforms \cite{colloquium,inscond}. Calculation of topological invariants allows prediction of edge states of different types (currents and localized modes) \cite{halperin,hatsugai}. Robustness of these edge states, particularly of topological currents, e.g. the quantum Hall state, is extraordinarily precise in its current measurement \cite{klitzing,measure1}. By now, we understand well the allowed topological states based on symmetry and dimension in Hermitian/ equilibrium systems \cite{weak1,schnyder1,twisted}, and we often know the appropriate topological invariants in these systems with even analytical calculation of them. However, as energy gain and loss are ubiquitous, it has become of interest to develop topological invariants for non-Hermitian systems \cite{38fold,gong,nhreview,nhinvariants}, yet existing results only address topological invariants for localized states, often in low-dimensional models.

The difficulty in analytically calculating topological invariants for currents in non-Hermitian systems is because such systems have at least two dimensions and more degrees of freedom than their Hermitian counterparts, which causes great difficulty in obtaining analytical expressions for their spectral properties. This precludes the study of the much richer properties that such higher dimensional systems can have compared to systems with only localized edge states. The lack of analytical descriptions is particularly troublesome in non-Hermitian systems due to the presence of common singularities such as exceptional points \cite{ep1,ep2,ep3}, which can be difficult to explore numerically. As the detection of topological properties typically relies on detecting phase jumps or singularities, the presence of analytical solutions would render this process much more accurate as compared to using numerics.

This lack of clean and precise analytical approaches contributes to gaps and confusing aspects in the literature. For instance, different eigenstate bases are often employed, from biorthogonal to right-only eigenstates, with it being unclear whether only one or both approaches will work \cite{leftright1,leftright2,leftright3,leftright4}. And while topological currents have been observed in non-Hermitian models, it remains unclear what their topological invariant is -- particularly as systems such as the one considered in this paper only admit weak topological invariants treating the Brillouin zone as a torus rather than the sphere. The development of analytical solutions for such systems would enable more accurate detection and understanding of these models, while permitting the development of newer mathematical approaches such as Morse theory. 

In this paper, we solve the full spectrum of a non-Hermitian topological model with currents \cite{tang1} and demonstrate its topological invariant analytically. This uses closed-form expressions for eigenstates and divisions in the phase diagram where different types of exceptional points exist. While a topological invariant had been previously proposed \cite{tang1} as an extension of the 2D Hermitian limit \cite{wakabayashi2} and investigated numerically \cite{numerical1}, there remained several confusing and ambiguous issues, such as a lack of numerical stability. Our approach allows the use of Morse theory to determine the precise topology of exceptional points in momentum space and show that they only form contractible loops within which the winding structure is unaffected, meaning that the topological invariant can be defined as an integration over the Brillouin zone minus exceptional points. This work also resolves the ambiguity about which eigenbases are usable. Lastly, we solve analytically for the number of edge modes in the Hermitian case (which was previously done only numerically \cite{wakabayashi2}), and provide an indication of how this can be extended to the non-Hermitian case. This work establishes new theoretical and mathematical tools which can be useful in the analysis of complex non-Hermitian systems more broadly.

% The 2D Zak phase captures the winding numbers of a function that describes the relative phase of eigenstate entries. In the trivial phase, this function varies smoothly over the Brillouin, whereas in the topological phase the function winds once in each of the $k_x$ and $k_y$ directions.

\begin{figure*}[t]
    \includegraphics[width=\linewidth]{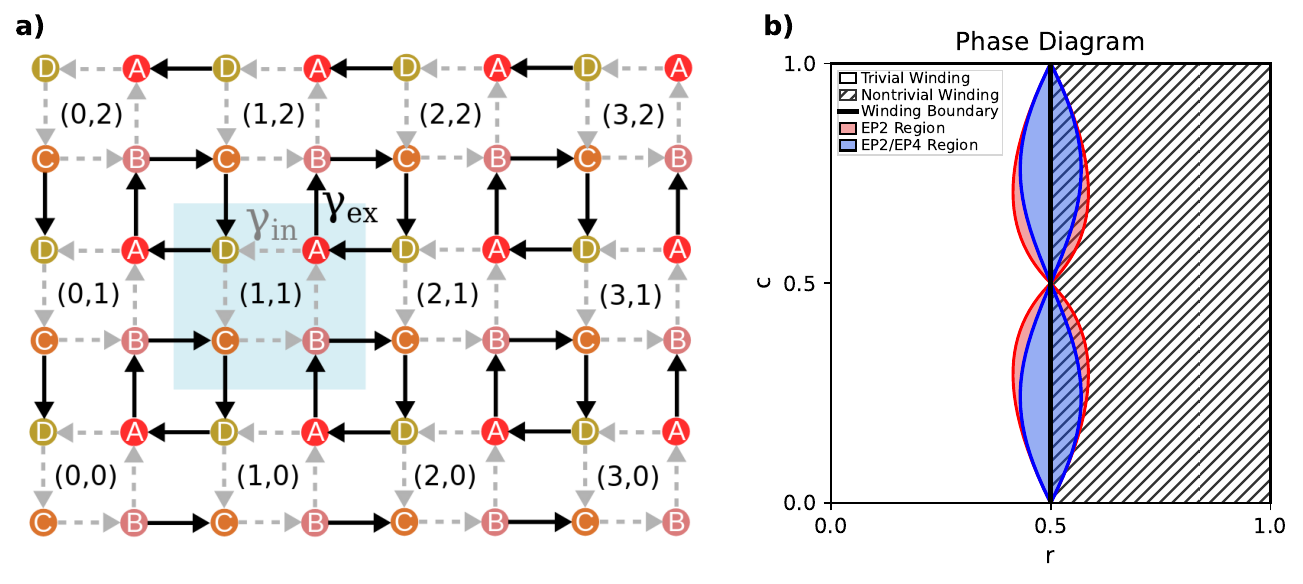}
    \caption{a) Representation of the physical model, with four sublattices and transition rates labelled. b) The phase diagram shows the topological regime for $r>\frac{1}{2}$ and the trivial regime for $r<\frac{1}{2}$. The red region marks where there exist $2^\text{nd}$ order exceptional points (EP2s) within the Brillouin zone and the blue region marks where there exist both $2^\text{nd}$ and $4^\text{th}$ order exceptional points (EP2s/EP4s) within the Brillouin zone.}
    \label{long}
\end{figure*}

\section{Symmetry and Relevant Topological Invariant}

We consider a 2d non-Hermitian model proposed by Tang et al. \cite{tang1}, which describes a stochastic system given by the master equation. This model can also describe a quantum tight-binding model with $4$ sublattices labelled $A,B,C,D$ with internal transitions $(x,y)_{A}\rightarrow(x,y)_B\rightarrow(x,y)_C\rightarrow (x,y)_D$ all at rate $\gamma_\text{in}$ and external transitions $(x,y)_A\rightarrow (x,y+1)_B\rightarrow (x+1,y+1)_C\rightarrow(x+1,y)_D\rightarrow(x,y)_A$ at rate $\gamma_\text{ex}$. This is shown schematically in Fig.~\ref{long}. Reverse transitions are then introduced in the opposite directions at rates $\gamma_\text{in}'$ and $\gamma_\text{ex}'$, respectively. Currents were observed in the stochastic system \cite{tang1} and the quantum version has identical properties, as they only differ by a diagonal term which is proportional to the identity (so the spectrum differs by a constant shift while the eigenstates and hence topological properties remain identical). 

Defining 
\begin{equation}
\begin{split}
    c(1-r) \equiv \gamma_\text{in}, \enspace  (1-c)(1-r) \equiv \gamma_\text{in}',&  \\
    cr \equiv \gamma_\text{ex}, \enspace  (1-c)r \equiv \gamma_\text{ex}',&
    \end{split}
\end{equation}
$c$ can be interpreted as the hermiticity of the system and $r$ can be interpreted as the chirality of the system.

The Bloch Hamiltonian is 
\begin{equation}
\resizebox{\columnwidth}{!}{$
H(k)=
    \begin{pmatrix}
    0 & \gamma_\text{in}+\gamma_\text{ex}' e^{-ik_y} & 0 & \gamma_\text{ex}e^{-ik_x} +\gamma_\text{in}'\\
    \gamma_\text{ex} e^{ik_y}+\gamma_\text{in}' & 0 & \gamma_\text{in}+\gamma_\text{ex}'e^{-ik_x} & 0 \\
    0 & \gamma_\text{ex} e^{ik_x} +\gamma_\text{in}'& 0 & \gamma_\text{in}+\gamma_\text{ex}'e^{ik_y} \\
    \gamma_\text{in}+\gamma_\text{ex}'e^{ik_x} & 0 & \gamma_\text{ex} e^{-ik_y} +\gamma_\text{in}'& 0 
    \end{pmatrix}_, \label{ham}
    $}
\end{equation} 
with the crystal (periodic) momentum $(k_x,k_y)$ taking values in $\mathbb{T}^2$.

Denoting
\begin{equation}
\begin{split}
        &A \equiv c(1-r)+(1-c)r e^{ik_x}, \enspace B \equiv c(1-r)+(1-c)r e^{ik_y}, \\
        &C \equiv (1-c)(1-r)+cr e^{ik_x},  \enspace D \equiv (1-c)(1-r)+cr e^{ik_y},
 \end{split}
 \label{defsabcd}
\end{equation}
allows the Hamiltonian to be written as
\begin{equation}
H(k)=
    \begin{pmatrix}
    0 & B^\ast  & 0 & C^\ast \\
    D & 0 & A^\ast  & 0 \\
    0 & C & 0 & B \\
    A & 0 & D^\ast  & 0 
    \end{pmatrix}_. 
\end{equation} 

\subsection{Real Momentum Space}
\label{secgbz}
The non-Hermitian skin effect is a uniquely non-Hermitian phenomenon associated with an accumulation of boundary states \cite{nhinvariants,skin1,skin2,skin3}. In \cite{2dskin}, the 2D non-Hermitian skin effect is discussed in detail and here their method is used to determine that the skin effect is suppressed in this system and why, therefore, both momentum $(k_x,k_y)$ can be taken to be real. The non-Hermitian skin effect comes from the considering the Generalized Brillouin zone, when $\beta_x \equiv^{ik_x}$, which does not necessarily obey $|\beta_x|=1$.
Consider the characteristic equation
\begin{equation}
    \text{det}[H(\beta_x,k_y)-E]=0, \label{skineq}
\end{equation}
where $k_y$ is real and $\beta_x$ is in general complex.

Fix a real $k_y$ and vary $\beta_x$. Number the solutions to this equation as $|\beta_{x,1}|\leq |\beta_{x,2}| \leq |\beta_{x,3}|\leq |\beta_{x,4}|$. Then the Generalized Brillouin Zone is given by varying $\beta_{x,2}$ and $\beta_{x,3}$ whilst imposing the condition $|\beta_{x,2}|=|\beta_{x,3}|$.

However, since Eq.~\ref{skineq} is invariant under $\beta_x \leftrightarrow \beta_x^{-1}$, solutions are paired reciprocally. Therefore, $|\beta_{x,1}=|\beta_{x,4}|^{-1}$ and $|\beta_{x,2}|=|\beta_{x,3}|^{-1}$. Therefore, when the Generalized Brillouin Zone condition $|\beta_{x,2}|=|\beta_{x,3}|$ is also imposed, $|\beta_{x,2}|=|\beta_{x,3}|=1$, reducing to the standard case of the usual Brillouin zone with both momenta real.

\subsection{Symmetry Classification}

There have been extensions of the tenfold way to non-Hermitian systems \cite{38fold,gong}. In their language, the Hamiltonian in this paper has time reversal symmetry since $\mathcal{T}^+H^\ast (k){\mathcal{T}^+}^{-1}=H(-k)$ for $\mathcal{T}^+= I_4$ (the $4d$ identity matrix). The superscript of $\mathcal{T}^+$ reflects that it squares to $+I_4$. The Hamiltonian also has sublattice symmetry since $\mathcal{S}_+H(k)\mathcal{S}_+^{-1}=-H(k)$ for $S \equiv I \otimes \sigma_z$, with the subscript denoting that $\mathcal{S}_+$ commutes with  $\mathcal{T}^+$.

This places the model in class $AI$ (time reversal symmetry which squares to $+1$) with $S_+$ (sublattice symmetry which commutes with time reversal symmetry), which admits no strong topological invariants in 2 dimensions. 

However, as mentioned in their same paper, they only considered strong invariants which treat the B.Z. as a sphere. When the B.Z. is a torus, weak lower-dimensional invariants can occur \cite{weak1,weak2}. For example, at a fixed $k_y$ slice, the system being considered has sublattice symmetry only, which is classified in 1 dimension as having a $\mathbb{Z} \oplus \mathbb{Z}$ point-gap invariant and $\mathbb{Z}$ line-gap invariant. 

Concretely, consider a slice of the Brillouin zone with fixed $k_y=\tilde{k_y}$. The induced 1D Hamiltonian is
$H(k_x,\tilde{k_y})$ which inherits the symmetry 
\begin{equation}
    \mathcal{S}H(k_x,\tilde{k_y})\mathcal{S}=-H(k_x,\tilde{k_y}),
\end{equation}
for $\mathcal{S}\equiv I_2 \otimes \sigma_z$. A 1D Hamiltonian with just this symmetry is class $A$ with $S$, which gives the point-gap invariant $\mathbb{Z}\oplus \mathbb{Z}$.

Except for $\mathcal{T}^+$-invariant lines in the Brillouin zone, the induced 1D Hamiltonian does not obey a time-reversal symmetry.

Without further considerations, the expected topological invariant would be $(\mathbb{Z}\oplus\mathbb{Z})_{k_x}\oplus(\mathbb{Z}\oplus \mathbb{Z})_{k_y}$. Calculating this for the actual model of the Hamiltonian in each direction, the invariant collapses to ${(\mathbb{Z}_2})_{k_x}\oplus{(\mathbb{Z}_2})_{k_y}$. 

The $U_{C_4}$-symmetry enforces the invariant in each direction is equal, i.e. the invariant is broken down to the diagonal subgroup $(\mathbb{Z}_2\oplus \mathbb{Z}_2)/\mathbb{Z}_2$. Sewing together 1D slices is then compatible with the invariant described in this paper.

The analysis and explicit expression for the topological invariant written in this paper agrees with the classification work previously done. However, topological classifications of generic Hamiltonians do not give a complete description of concrete models, nor do they give a complete understanding of calculating explicit invariants. This paper does both for this specific model. 

This Hamiltonian also possesses two less standard symmetries.
Firstly, $PH(k_x,k_y)P^{-1}=H^\dagger(k_y,k_x)$ for
\begin{equation}
P \equiv
\begin{pmatrix}
1 & 0 & 0 & 0 \\[6pt]
0 & 0 & 0 & 1 \\[6pt]
0 & 0 & 1 & 0 \\[6pt]
0 & 1 & 0 & 0 
\end{pmatrix}. \label{gphsym}
\end{equation}
This is an order-two symmetry \cite{ordertwo1,ordertwo2,ordertwo3} in the sense that applying this transformation twice is the identity.

This Hamiltonian also obeys $U_{C_4}H(k_x,k_y)U_{C_4}^{-1}=H(-k_y,k_x)$ for
\begin{equation}
U_{C_4} \equiv
\begin{pmatrix}
0 & 1 & 0 & 0 \\[6pt]
0 & 0 & 1 & 0 \\[6pt]
0 & 0 & 0 & 1 \\[6pt]
1 & 0 & 0 & 0
\end{pmatrix}.
\end{equation}

This is an order-four symmetry in the sense that applying this transformation four times is the identity.

\subsection{Energies}
The energies of Eq.~\ref{ham} are
\begin{equation}
    E=\pm \sqrt{a \pm \sqrt {a^2-b}}
\end{equation}
where
\begin{equation}
\begin{split} 
    a & \equiv r(1-r)[c^2+(1-c)^2]\left[\cos(k_x)+\cos(k_y)\right]\\
    &+2c(1-c)[r^2+(1-r)^2], \\
    b & \equiv 4r^2(1-r)^2c^2(1-c)^2\left[\cos(k_x)-\cos(k_y)\right]^2\\
    &-\left[r^2-(1-r)^2\right]^2 \left[c^2-(1-c)^2\right]^2.
\end{split} \label{ab}
\end{equation}

Since it an important quantity, make the definition
\begin{equation}
    \Phi \equiv a^2-b. \label{phi}
\end{equation}

Since the sign of $\Phi$ determines whether there are a pair of purely imaginary and purely real energies or there are two complex conjugate pairs of energies, it is useful to calculate whether there is a negative minimum of $\Phi$ over the Brillouin zone. It is proven in App.~\ref{app1} that there is only one minimum that can be negative. This same minimum lies at 4 points $(\pi, \pm u)$ and $(\pm u, \pi)$ within the Brillouin zone for
\begin{equation}
u \equiv \arccos \left[1-\frac{2c(1-c)[2c^2-2c+2r^2-2r+1]}{r(1-r)[c^2-(1-c)^2]^2}\right].
\end{equation}
This minimum is zero iff
\begin{equation}
\begin{split}
&2c(1-c)(2c^2-2c+2r^2-2r+1)\\
&-|(2r-1)(2c-1)|[c^2+(1-c)^2]=0.
\label{boundary}
\end{split}
\end{equation}
This can be rewritten for $r>\frac{1}{2},c>\frac{1}{2}$ as
\begin{equation}
    r=\frac{-2c^2+4c-1}{2c}. \label{firstgraph}
\end{equation}
For general $r$ and $c$, this expression becomes
\begin{equation}
    \left|r-\frac{1}{2}\right|=\left|\frac{\left|1-2c\right|\left(\left|1-2c\right|-1\right)}{2\left(\left|1-2c\right|+1\right)}\right|,
\end{equation}
which is shown in Fig.~\ref{long} as the dividing line between the presence and absence of exceptional points in the Brillouin zone. Determining whether $\Phi$ has a positive or negative global minimum in the Brillouin zone is equivalent to asking whether there exist $2^\text{nd}$ order exceptional points. The structure of these exceptional points will be explored in more detail in Sec.~\ref{sectopep}.

\subsection{Eigenstates}
Using the definitions in  Eqs.~\ref{defsabcd}, the eigenstate equations can be written as
\begin{equation}
\begin{split}
        &B^\ast  z_2 + C^\ast  z_4 = E z_1, \quad
        Dz_1 + A^\ast  z_3 = E z_2, \quad \\
        &Cz_2 + B z_4 = E z_3, \quad
        A z_1 + D^\ast  z_3 = E z_4.
\end{split} \label{rotations}
\end{equation}
These imply
\begin{equation}
    \begin{split}
        &\tau \equiv \frac{z_3}{z_1}=\frac{E^2-B^\ast  D-AC^\ast }{A^\ast B^\ast  + C^\ast D^\ast }=\frac{AB+CD}{E^2-A^\ast C-BD^\ast },\\
        &\frac{z_4}{z_2}=\frac{E^2-B^\ast D-A^\ast C}{A^\ast  B+C^\ast D}=\frac{AB^\ast +CD^\ast }{E^2-BD^\ast -AC^\ast }. \label{oh1}
    \end{split}
\end{equation}
Define a $\frac{\pi}{2}$ clockwise rotation in momentum space as $\mathcal{R} f(k_x,k_y)=f(-k_y,k_x)$, where $f(k_x,k_y)$ is some function (not necessarily continuous). This rotation transforms $A\rightarrow B^\ast , B \rightarrow A, C\rightarrow D^\ast , D\rightarrow C$ and $E \rightarrow E$.
Dividing through each of Eqs.\ref{rotations} by $z_2$ or $z_3$ results in
\begin{equation}
    \begin{split}
        &C^\ast \left(\frac{z_4}{z_2}\right)+B^\ast =E\left(\frac{z_1}{z_2}\right), \quad
        D\left(\frac{z_1}{z_3}\right)+A^\ast =E\left(\frac{z_2}{z_3}\right),\\
        &C\left(\frac{z_2}{z_4}\right)+B=E\left(\frac{z_3}{z_4}\right), \quad
        D^\ast \left(\frac{z_3}{z_1}\right)+A=E\left(\frac{z_4}{z_1}\right).
    \end{split}
\end{equation}
$\mathcal{R}$ applied to each LHS gives the next LHS (applied to the fourth also gives the first), so
\begin{equation}
\left(\frac{z_1}{z_2}\right)=\mathcal{R}\left(\frac{z_4}{z_1}\right)=\mathcal{R}^2\left(\frac{z_3}{z_4}\right)=\mathcal{R}^3\left(\frac{z_2}{z_3}\right)=\mathcal{R}^4\left(\frac{z_1}{z_2}\right).\label{oh3}
\end{equation}
Without loss of generality, taking $z_1$ to be $1$ and not yet specifying $z_2$ implies the eigenstate is of the form
\begin{equation}
    \frac{1}{N_1}\begin{pmatrix} 
     1 \\
    z_2 \\
z_2 \mathcal{R}(z_2) \\
     z_2 \mathcal{R}(z_2)\mathcal{R}^2(z_2)
    \end{pmatrix} , \label{eig10}
\end{equation}
where
\begin{equation}
    z_2 \mathcal{R}(z_2) \mathcal{R}^2(z_2) \mathcal{R}^3(z_2)=1, \label{additional}
\end{equation}
and $N_1$ is some overall normalization constant.

Using $\tau$ defined in Eq.~\ref{oh1} and the second equation of Eqs.~\ref{rotations} means any eigenstate can be written as
\begin{equation}
    \frac{1}{N_1}\begin{pmatrix}
     1 \\
    \left( D+A^\ast  \tau \right)/E \\
\tau \\
     \left( A+D^\ast  \tau \right)/E
    \end{pmatrix}. \label{eig11}
\end{equation}

The expressions in Eq.~\ref{eig10} and Eq.~\ref{eig11} are the same under the identifications
\begin{equation}
        z_2 \equiv \frac{D+A^\ast \tau}{E}, \quad 
        \mathcal{R}(z_2) \equiv \frac{E\tau}{D+A^\ast \tau}.
\end{equation}

Using Eq.~\ref{oh1} then implies
\begin{equation}
    \mathcal{R}(\tau) = \frac{A\tau^{-\frac{1}{2}}+D^\ast \tau^{\frac{1}{2}}}{D\tau^{-\frac{1}{2}}+A^\ast \tau^{\frac{1}{2}}}. \label{rot90}
\end{equation}

This motivates the definition
\begin{equation}
    \mu \equiv  A\tau^{-\frac{1}{2}}+D^\ast  \tau^{\frac{1}{2}}. \label{mudef}
\end{equation}

For convenience, denote the $\frac{\pi}{2}$ rotation action of any function $f$ by $f'$.

Then an eigenstate can be written as
\begin{equation}
    \frac{1}{N_2}\begin{pmatrix}
    \mu'^{3/4} \mu'''^{-1/4}\\
    \pm \mu''^{3/4}\mu^{-1/4} \\
\mu'''^{3/4}\mu'^{-1/4} \\
    \pm  \mu^{3/4}\mu''^{-1/4}
    \end{pmatrix} . 
\end{equation}
with $\mu \mu' \mu'' \mu'''= E^4$. This describes four eigenstates, one from the choice of sign here and one from the choice of sign of $E^2$ in the definition of $\tau$.

Finally, the normalized eigenstate can be written as
\begin{equation}
|R(k_x,k_y)\rangle =\begin{pmatrix}
    re^{i\theta}\\
    r'e^{i\theta'} \\
r''e^{i\theta''} \\
     r'''e^{i\theta'''}
    \end{pmatrix}.  \label{trivgauge}
\end{equation}
where
\begin{equation}
\begin{split}
    r &\equiv \frac{\left|\mu'^{3/4}\mu'''^{-1/4}\right|}{\sqrt{
    \begin{split}&\left|\mu'^{3/4}\mu'''^{-1/4}\right|^2+\left|\mu''^{3/4}\mu^{-1/4}\right|^2\\
    +&\left|\mu'''^{3/4}\mu'^{-1/4}\right|^2+\left|\mu^{3/4}\mu''^{-1/4}\right|^2
    \end{split}}},
    \end{split} \label{rdef}
\end{equation}
and
\begin{equation}
    \theta \equiv \text{phase}(\mu'^{3/4}\mu'''^{-1/4}). \label{thetadef}
\end{equation} 

Note that $r$ has no poles or zeros in the Brillouin zone, except at $4^\text{th}$ order exceptional points, when either $r=r''=0$ and $r'=r'''=\frac{1}{\sqrt{2}}$ or $r'=r'''=0$ and $r=r''=\frac{1}{\sqrt{2}}$. This case is dealt with separately in Sec.~\ref{secep4}, since there is no consistent choice of eigenstate that can avoid such points. Away from these points, $r(k_x,k_y)$ is nonzero and not singular. This will be important because all the nontrivial structure of the eigenstates will be contained in the phases of entries, which are well-defined when $r$ has no zeros or poles. 

There is also the case that $r$ is badly defined for two out of four states when $b=0$ (defined in Eq.~\ref{ab}) somewhere in the Brillouin zone. However, this can be avoided by simply considering the two states that avoid $b=0$ everywhere in the Brillouin zone.

\subsection{Topological Invariant}
In this section, a discussion of why the 2D Zak phase is the right invariant to consider is given, as well as a discussion of which eigenstate basis and normalization to use. It will be proven that the 2D Zak phase captures the winding numbers of the function $2\theta$ defined by Eq.~\ref{thetadef}.

It was shown in \cite{wakabayashi2} that the topological invariant for the Hermitian limit of this model is the 2D Zak phase $(Z_{k_i},Z_{k_y})$ defined as
\begin{equation}
    Z_{k_i}\equiv \frac{1}{2\pi} \int_{\text{B.Z.}} d^2 k \frac{\langle L | -i \partial_{k_i}|R\rangle}{\langle L|R\rangle}. \label{lrzak}
\end{equation}

It was suggested in \cite{tang1} that this invariant would extend to the non-Hermitian regime. It will be shown in this paper that this is indeed also the right topological invariant for the non-Hermitian case. An important point to highlight is that this should be calculated for left and right eigenstates smoothly defined over the whole Brillouin zone. 

It will be shown that an appropriate gauge choice for these eigenstates is
\begin{equation}
\begin{split}
|R(k_x,k_y)\rangle=&e^{-i\theta''}
\begin{pmatrix}
    re^{i\theta}\\
    r'e^{i\theta'} \\
r'' e^{i\theta''}\\
     r'''e^{i\theta'''}
    \end{pmatrix}=
\begin{pmatrix}
    re^{i(\theta-\theta'')}\\
    r'e^{i(\theta'-\theta'')} \\
r'' \\
     r'''e^{i(\theta'''-\theta'')}
    \end{pmatrix},  \\
|L(k_x,k_y)\rangle=&e^{-i\theta''}P|R(-k_y,-k_x)\rangle\\
=& e^{-i\theta''}
\begin{pmatrix}
    r(-k_y,-k_x)e^{i\theta(-k_y,-k_x)}\\
    r(-k_x,k_y)e^{i\theta(-k_x,k_y)} \\
r(k_y,k_x) e^{i\theta(k_y,k_x)}\\
     r(k_x,-k_y)e^{i\theta(k_x,-k_y)}
    \end{pmatrix}.
    \end{split} \label{topgauge}
\end{equation}

Note that $Z_{k_i}$ is only defined modulo $2\pi$. It is possible to make the $U(1)$ gauge transformation
\begin{equation}
\begin{split}
&|R(k_x,k_y)\rangle \rightarrow e^{i\phi(k_x,k_y)}|R(k_x,k_y)\rangle,\\ \enspace \langle &L(k_x,k_y)| \rightarrow \langle L(k_x,k_y| e^{-i\phi(k_x,k_y)},
\end{split}
\end{equation}
if $e^{i\phi(k_x,k_y)}$ is smooth over the entire Brillouin zone. Global smoothness of $e^{i\phi(k_x,k_y)}$ is equivalent to $\phi(k_x,k_y)$ obeying the boundary conditions 
\begin{equation}
\begin{split}
    &\phi(k_x+2\pi,k_y)=\phi(k_x,k_y)+2\pi m , \\
    &\phi(k_x,k_y+2\pi)=\phi(k_x,k_y)+2\pi n,
\end{split}
\end{equation}
for winding numbers $(m,n) \in \mathbb{Z}^2$.

Therefore, $Z_{k_i}$ should be invariant under such a gauge transformation, which transforms
\begin{equation}
    Z_{k_i} \rightarrow Z_{k_i} + \frac{1}{2\pi} \int_{\text{B.Z.}} d^2k \enspace  (\partial_{k_i} \phi).
\end{equation}
This is equivalent to
\begin{equation}
    (Z_{k_x},Z_{k_y}) \rightarrow (Z_{k_x},Z_{k_y})+2\pi(m,n).
\end{equation}
Therefore, $Z_{k_i}$ are defined modulo $2\pi$.

Define
\begin{equation}
    f(k_x,k_y)\equiv r(k_x,k_y)e^{i\theta(k_x,k_y)}.
\end{equation}
Calculating the Zak phase for the gauge choice in Eq.~\ref{topgauge} gives
\begin{equation}
\begin{split}
    Z_{k_i}
    = &\frac{1}{2\pi}\int  d^2 k\enspace [f^\ast (-k_y,-k_x) \partial_{k_i} f(k_x,k_y)\\
    + &f^\ast (-k_x,k_y) \partial_{k_i}f(-k_y,k_x)\\
    +&f^\ast (k_y,k_x) \partial_{k_i}f(-k_x,-k_y)\\
    + &f^\ast (k_x,-k_y) \partial_{k_i}f(k_y,-k_x)]/(\langle L | R \rangle)\\
    -&\frac{1}{2\pi}\int  d^2 k \enspace\frac{1}{2}\partial_{k_i} (2\theta(-k_x,-k_y)).
    \end{split} \label{cancelling}
\end{equation}

The first and third term cancel by a change of variables $(k_x,k_y)\rightarrow (-k_x,-k_y)$ since $\langle L| R \rangle$ is invariant under this transformation. Similarly, the second and fourth term cancel. This leaves the last term which can be written as
\begin{equation}
    Z_{k_i}=\frac{1}{4\pi} \int_\text{B.Z.} d^2 k \enspace  \partial_{k_i} (2\theta(k_x,k_y)) = \pi n_i. \label{linkwinding}
\end{equation}
where $n_x$ and $n_y$ are the winding numbers of $2\theta$. Therefore, the Zak phase in each direction is simply proportional to the winding number of $2\theta$ in each direction. In this way, the topological invariant of this system in completely determined by a single function $2\theta(k_x,k_y)$. 

Note that an identical calculation leads to the same result for the Zak phase defined with the position of left and right eigenstates swapped.

It is shown in App.~\ref{symmetriesapp} that the symmetries of this model constrain the Zak phase. For this particular model, the $C_4$-symmetry implies that $Z_{k_x}=Z_{k_y}$. 

\subsection{Choice of eigenstate basis}

This particular model makes it more convenient to calculate the Zak phase using only right eigenstates as parameterized in Eq.~\ref{topgauge} since
\begin{equation}
\begin{split}
    \frac{1}{2\pi} \int_\text{B.Z.} d^2k &\enspace \langle R| -i\partial{k_i}| R\rangle\\
    =\frac{1}{2\pi} \int d^2k &\enspace [f^\ast(k_x,k_y)\partial_{k_i}f(k_x,k_y)\\
    &+f^\ast(-k_y,k_x) \partial_{k_i} f(-k_y,k_x)\\
    &+f^\ast(-k_x,-k_y) \partial_{k_i} f(-k_x,-k_y)\\
    &+f^\ast(k_y,-k_x) \partial_{k_i} f(k_y,-k_x)]\\
    -&\frac{1}{2\pi}\int  d^2 k \enspace\frac{1}{2}\partial_{k_i} (2\theta(-k_x,-k_y)).
\end{split}
\end{equation}
Similarly to Eq.~\ref{cancelling}, the first and third terms cancel, as well as the second and fourth. This leaves the same result
\begin{equation}
    Z_{k_i}=\frac{1}{4\pi} \int_\text{B.Z.} d^2 k \enspace  \partial_{k_i} (2\theta(k_x,k_y)) = \pi n_i.
\end{equation}
Therefore, this definition of the Zak phase using only right eigenstates is equivalent for this particular model. They will be used interchangeably, although the issue of normalizing states will be discussed for the case when exceptional points are present.

Note that, using the normalization $\langle R|R \rangle=1$ gives a real Zak phase since
\begin{equation}
\begin{split}
    Z_{k_i}^\ast=&\int_\text{B.Z.} d^2k \enspace \langle R |^\ast i\partial_{k_i}|R\rangle^\ast\\
    =&\int_\text{B.Z.} d^2k \enspace  i\partial_{k_i} \left(\langle R|R \rangle\right)  +\langle R |-i\partial_{k_i}|R\rangle\\
    =&Z_{k_i}.
\end{split}    
\end{equation}

It is shown, using sublattice and time-reversal symmetry, in App.~\ref{appband} that calculating the 2D Zak phase gives the same answer when using each eigenstate of the $4$ energy bands.

It is also shown in App.~\ref{symmetriesapp} that left and right eigenstates can be swapped in the definition of the Zak phase, due to the symmetry in Eq.~\ref{gphsym} applied to this particular model.

\section{Region Without Exceptional Points}

Here, the case with no exceptional points is considered, as shown in Fig.~\ref{long}.b. Since calculating the 2D Zak phase is equivalent to calculating the winding number of the function $2\theta(k_,k_y)$, it will be proven that the winding numbers are $(0,0)$ for $r<\frac{1}{2}$ and $(1,1)$ for $r>\frac{1}{2}$. Furthermore, it will be shown that the gauge choice in Eq.~\ref{topgauge} is well-defined across the whole Brillouin zone.

Since it is a result that will be used multiple times in this paper, it is good to highlight the following. For a smooth real function on the Brillouin zone, say $\Pi(k_x,k_y)$, if there are no points for which $\Pi=c$ and $\nabla \Pi=0$, then the level sets $\Phi^{-1}(c)$ are properly embedded 1D submanifolds of the Brillouin zone (see, for example, Corollary 5.14 of \cite{lee03}). This in turn implies that the level sets $\Phi^{-1}(c)$ are the disjoint union of simple closed curves. 

\subsection{Eigenstate Simplification}
When there are no exceptional points, $E$ is real or purely imaginary. In this case, the form of eigenstates is significantly simplified.

Eq.~\ref{oh1} implies that $\tau^\ast =\tau^{-1}$ which in turn implies $|\tau|=1$. 

This also implies that for $\mu$ defined in Eq.~\ref{mudef}, $\mu(-k_x,-k_y)=\mu^\ast(k_x,k_y)$. Eqs.~\ref{rdef} and \ref{thetadef} then imply $r(k_x,k_y)=r(-k_x,-k_y)$ and $\theta(k_x,k_y)=-\theta(-k_x,-k_y)$.

This means that the eigenstates in Eq.~\ref{topgauge} simplify to
\begin{equation}
    |R(k_x,k_y)\rangle=\begin{pmatrix}
      re^{2i\theta} \\
    r' e^{i(\theta+\theta')} \\
 r \\
     r' e^{i(\theta-\theta')}
    \end{pmatrix} . \label{eig1}
\end{equation}

Note that $r$ never has zeros or poles except when $E=0$.

$\tau\equiv e^{-2i\theta}$ can be written as 
% \begin{equation}
% \begin{split}
% \tau & =\frac{\pm \sqrt{\Phi}+ir(1-r)(c^2-(1-c)^2)(\sin(k_x) - \sin(k_y))}{(c^2+(1-c)^2)\left[(1-r)^2+r^2e^{-ik_x-ik_y}\right]+2r(1-r)c(1-c)(e^{-ik_x}+e^{-ik_y})} \\
% &= \frac{(c^2+(1-c)^2)\left[(1-r)^2+r^2e^{ik_x+ik_y}\right]+2r(1-r)c(1-c)(e^{ik_x}+e^{ik_y})} {\pm \sqrt{\Phi}-ir(1-r)(c^2-(1-c)^2)(\sin(k_x) - \sin(k_y))}. \label{tau}
% \end{split}
% \end{equation}
% Splitting the numerator and denominator into real and imaginary parts, Eqs.~\ref{tau} are
\begin{equation}
        \tau= \frac{F+iG}{H-iJ}=\frac{H+iJ}{F-iG} \label{taufrac}
\end{equation}
where $F,G,H,J$ are the real functions
\begin{equation}
\begin{split}
F  \equiv & \pm \sqrt{\Phi},\\
G \equiv & r(1-r)(c^2-(1-c)^2)(\sin(k_x)-\sin(k_y)),\\
H\equiv & (1-r)^2[c^2+(1-c)^2]\\
&+2rc(1-r)(1-c)[\cos(k_x)+\cos(k_y)]\\
& +r^2[c^2+(1-c)^2]\cos(k_x+k_y),\\
J \equiv & r^2\left[c^2+(1-c)^2\right] \sin(k_x+k_y)\\
&+2rc(1-r)(1-c)(\sin(k_x)+\sin(k_y)).
\end{split}
\end{equation}
In this notation, the identity
\begin{equation}
    F^2+G^2=H^2+J^2. \label{squares}
\end{equation}
holds, which directly verifies that $\tau$ has magnitude 1. 

\subsection{Calculation of Winding Number}

As was shown in Eq.~\ref{linkwinding}, the Zak phase calculates the nontrivial winding of $2\theta(k_x,k_y)$ in each direction. Given that $\tau \equiv e^{-2i\theta}$, if there is nontrivial winding of $2\theta$, $2\theta$ must go through the values $-\pi$ and $\pi$ (without loss of generality, choosing $2\theta=-\pi$ as the lowest value of $2\theta$ corresponding to $\tau=-1$). 

Since it has been argued that all energy bands give the same Zak phase, without loss of generality, consider $F=\pm \sqrt{\Phi}$ to have the positive choice of sign. 

Also since $\tau$ is invariant under the simultaneous transformations $k_x \leftrightarrow k_y$ and $c\leftrightarrow(1-c)$, without loss of generality, consider $c>\frac{1}{2}$. 

\begin{figure}
    \centering
    \includegraphics[width=\columnwidth]{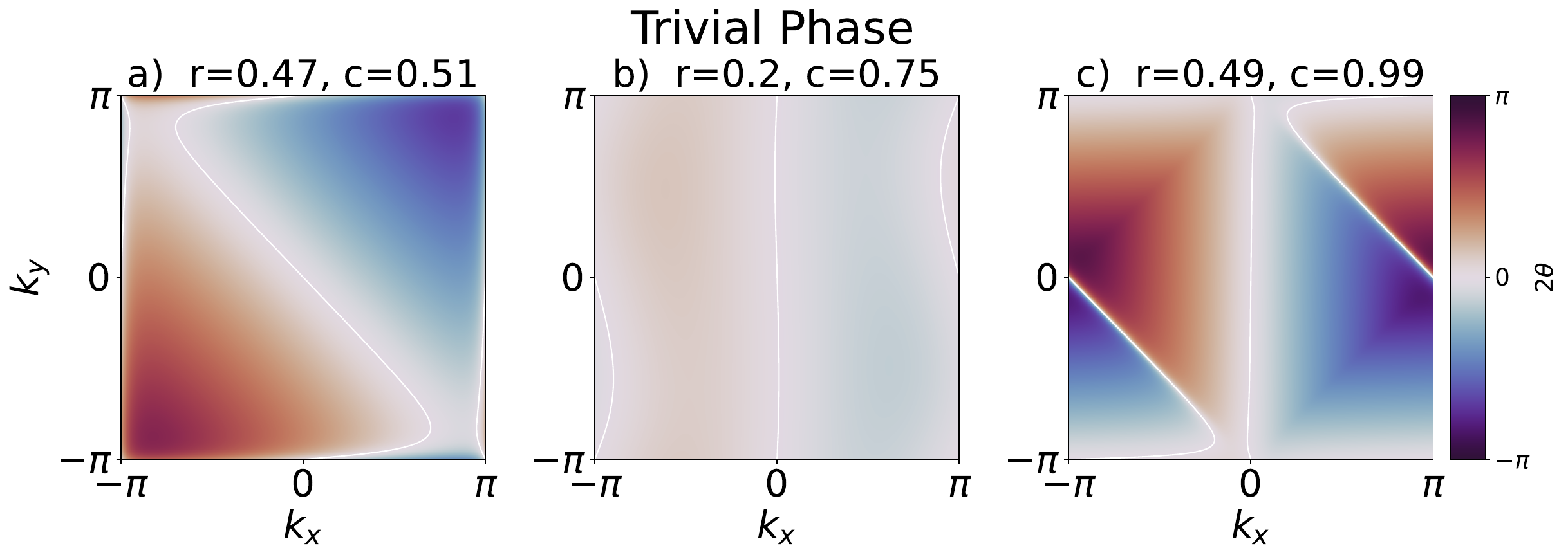}
    \caption{These figures show the function $2\theta$ defining the phase of eigenstate entries as in Eq.~\ref{topgauge} globally defined over the Brillouin zone without any discontinuities, for $(r,c)$ values in the trivial regime without exceptional points. Zero winding numbers of $2\theta$ lead to zero Zak phase. Zero contours of $\theta$ are highlighted with white lines for clarity.}
    \label{Fig1}
\end{figure}
From Eq.~\ref{taufrac}, $\tau=-1$ is equivalent to $G-J=0$ and $H<0$ whereas $\tau=1$ is equivalent to $G+J=0$ and $H>0$. It is proven in App.~\ref{g} that $G-J$ is never zero with zero gradient in the Brillouin zone which implies zero sets of $G-J$ are a union of disjoint closed curves. It is also shown in App.~\ref{g} that there are no contractible loops on which $G-J=0$ in the Brillouin zone.

Any noncontractible loop along which $G-J=0$ must pass through the line $k_x+k_y=0$. Along this line, for all $r$, $G-J=0$ only at $(0,0)$ and $(0,\pi)$, which implies that there are always exactly two lines of zeros of $G-J$. $H(0,0)>0$ and $H$ cannot change sign on a zero set of $G-J$, so $H>0$ along the entire zero set of $G-J$ passing through $(0,0)$. 

$H(0,\pi)=(1-2r)[c^2+(1-c)^2]$. Therefore, for $r<\frac{1}{2}$, $H>0$ along both zero sets of $G-J$ and, therefore, $\tau$ never takes the value $-1$ anywhere in the Brillouin zone. This implies that $2\theta$ is smoothly defined over the entire Brillouin zone with zero winding number. This also implies that $(\theta+\theta')$ and $(\theta-\theta')$ are also smoothly defined over the Brillouin zone, meaning that the eigenstates written in Eq.~\ref{topgauge} are well-defined and the Zak phases are 
\begin{equation}
    (Z_{k_x},Z_{k_y})=(0,0).
\end{equation}
Graphs showing $2\theta(k_x,k_y)$ smoothly defined over the whole Brillouin zone in the trivial phase are shown in Fig.~\ref{Fig1}.

On the other hand, since $H(0,\pi)<0$ for $r>\frac{1}{2}$, there is a unique noncontractible loop along which $G-J=0$ and $H<0$ which is equivalent to $\tau=-1$. There is also a unique loop along which $\tau=+1$, since there is a bijection between loops for which $G+J=0$ and loops for which $G-J=0$ given by $(k_x,k_y)\leftrightarrow(k_y,k_x)$. Since this map leaves $H$ invariant, the loop with $G-J=0$ and $H>0$ through $(0,0)$ is mapped to the loop with $G+J=0$ and $H>0$, which is the unique loop for which $\tau=+1$. Therefore, for $r>\frac{1}{2}$, there is one unique noncontractible loop for each of $\tau=+1$ and $\tau=-1$ in the Brillouin zone. They do not intersect since $H$ has opposite sign on each of them. These facts imply that $\tau=e^{-2i\theta}$ winds once from $-1$ to $+1$ and back to $+1$ in each of the momentum directions, which is equivalent to $2\theta$ winding once from  $-\pi$ to $+\pi$.

All that remains to show is that $(\theta+\theta')$ and $(\theta-\theta')$ only have $2\pi$-discontinuities in the Brillouin zone for the eigenstates to be smooth globally. $2\theta(k_x,k_y)$ has a $2\pi$-discontinuity from left to right along a line from $(0,\pi)$ to $(\pi,0)$ in the upper right quadrant of the Brillouin zone. Since $2\theta(-k_x,-k_y)=-2\theta(k_x,k_y)$, there is a similar $2\pi$-discontinuity in $2\theta$ from left to right in a line connecting $(-\pi,0)$ to $(0,-\pi)$ in the lower left quadrant of the Brillouin zone. This structure can be seen in Fig.~\ref{Fig42}.a. Adding $2\theta$ to $2\theta'$ gives a function $2(\theta+\theta')$ with winding numbers $(2,0)$ with two lines of $2\pi$-discontinuities in $2(\theta+\theta')$ joining $(0,-\pi)$ and $(0,\pi)$, one lying in the half of the Brillouin zone and one lying in the right half of the Brillouin zone. This implies that $(\theta+\theta')$ has winding numbers $(1,0)$ with a single line of $2\pi$-discontinuities, as shown in Fig.~\ref{Fig42}.b. A similar argument gives a well-defined $(\theta-\theta')$ with winding numbers $(0,1)$, as shown in Fig.~\ref{Fig42}.c.

Therefore, the eigenstates as written in Eq.~\ref{topgauge} are globally smooth and the Zak phase for $r>\frac{1}{2}$ due to the nontrivial winding of $2\theta$ is 
\begin{equation}
    (Z_{k_x},Z_{k_y})=(\pi,\pi).
\end{equation}

\begin{figure}
\centering\includegraphics[width=\columnwidth]{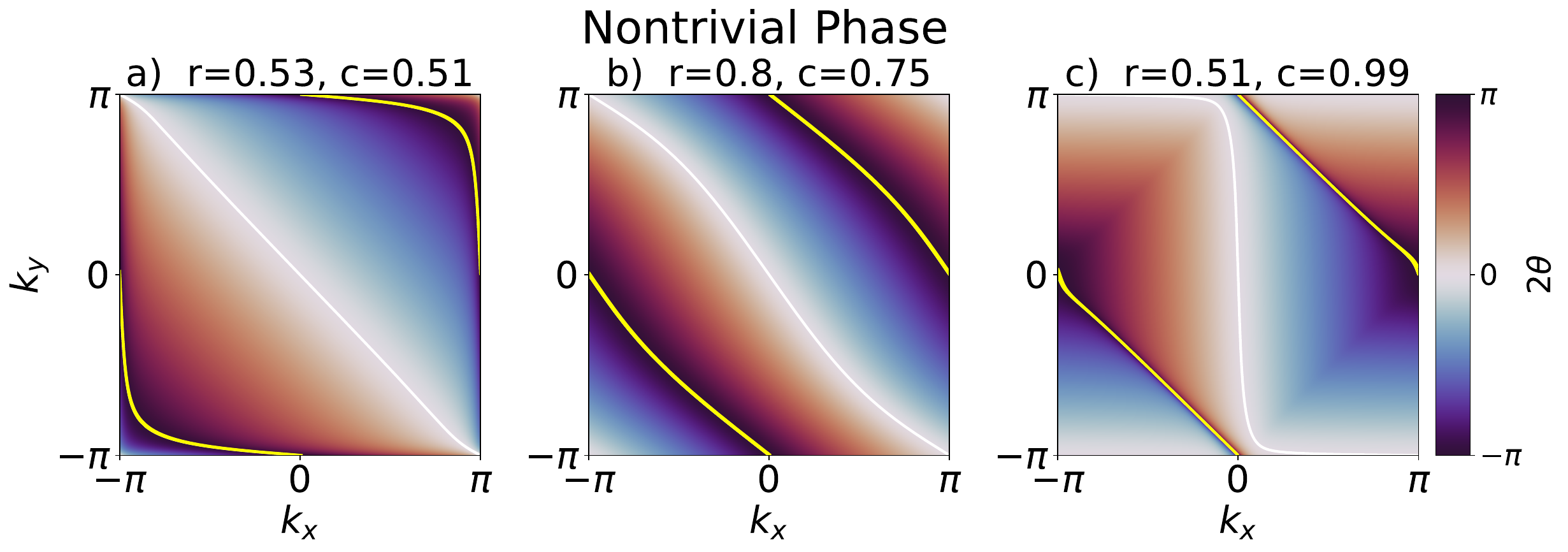}
\caption{These figures show the function $2\theta$ with nontrivial winding numbers in the topological regime without exceptional points. A $2\pi$-discontinuity is highlighted by yellow lines and zero sets of $2\theta$ are highlighted by white lines for clarity. Note that for each $k_x$ slice and $k_y$ slice, the discontinuity in $2\theta$ is $2\pi$. This gives rise to nontrivial winding numbers and therefore a nontrivial Zak phase in each direction.}
\label{Fig41}
\end{figure}
\begin{figure}
\centering\includegraphics[width=\columnwidth]{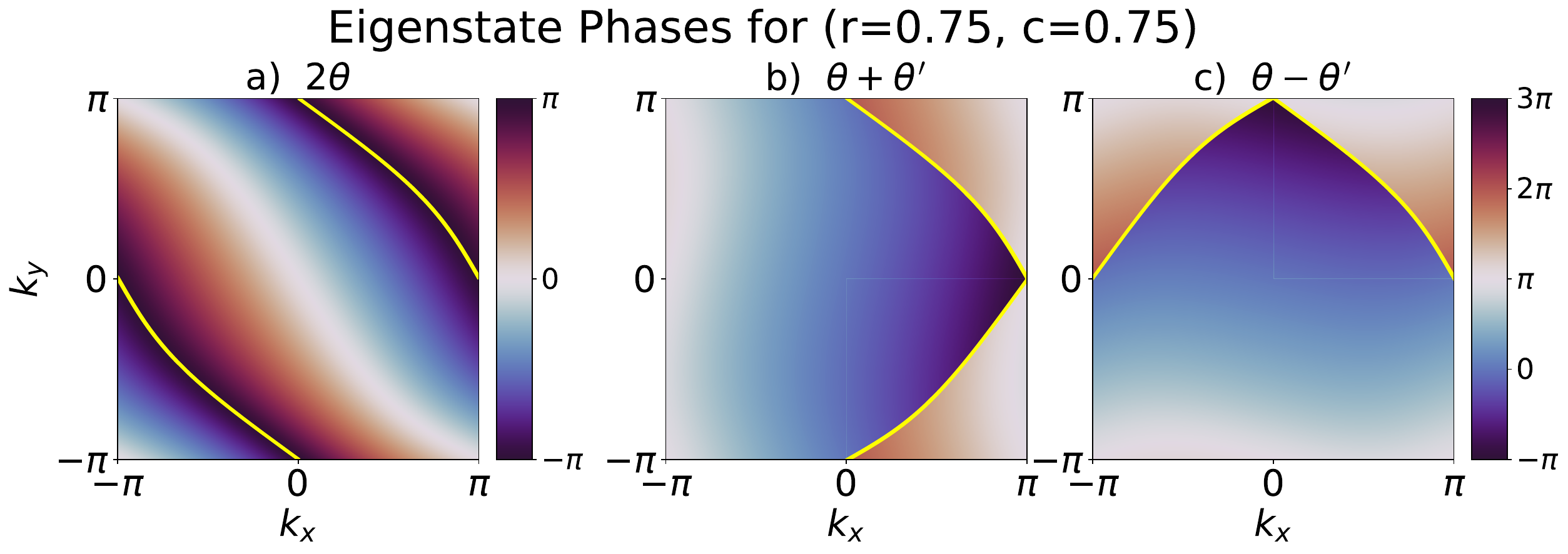}
\caption{These figures show the phases of different entries of the eigenstates written in Eq.~\ref{topgauge} for $(r,c)$ in the topological regime without exceptional points, showing that the phases only have discontinuities of multiples of $2\pi$ and therefore give smooth eigenstates.}
\label{Fig42}
\end{figure}

\subsection{Hermitian Limit of \texorpdfstring{2D}{2D} Zak Phase}
Previous work \cite{wakabayashi2} has established the eigenstates of the Hermitian system as
\begin{equation}
    \begin{pmatrix}
     -1 \\
   s_1e^{e^{-i\phi_x(k_x)}} \\
    s_2 e^{-i\phi_y(k_y)} \\
    -s_1 s_2 e^{-i(\phi_x(k_x)+\phi_y(k_y))} 
    \end{pmatrix} .
\end{equation}
where $s_1,s_2$ are $\pm 1$ and $\phi_x(k_x)$ and $\phi_y(k_y)$ are functions of only $k_x$ and $k_y$, respectively. This is the same as Eq.~\ref{topgauge} up to a reordering of entries since there is a split of $\theta(k_x,k_y)$ into $k_x$ and $k_y$ dependencies so the following identifications can be made: $\theta(k_x,k_y) \equiv - \frac{1}{2} \left(\phi_x(k_x) + \phi_y(k_y)\right)$ and $|r(k_x,k_y)|\equiv \frac{1}{2}$. Note that in the Hermitian limit $c=\frac{1}{2}$, $\phi_i(k_i)$ is a function such that $\phi_i(-k_i)=-\phi_i(k_i)$ for $i=x,y$.

\section{Region with Exceptional Points}
\label{sectopep}
The purpose of this section is to determine the topology of all types of exceptional points within the Brillouin zone then to comment on how they fit into the framework of the 2D Zak phase. There are two types of exceptional points for this model, $2^\text{nd}$ and $4^\text{th}$ order.

When $\Phi<0$ ($\Phi$ was defined in Eqs.~\ref{ab},\ref{phi}) somewhere within the Brillouin zone, there exist $2^\text{nd}$ order exceptional points where $\Phi=0$. When $\Phi=0$, $E=\pm \sqrt{a}$ and two pairs of eigenstates coalesce into two eigenstates. 

When $a=b=0$, there is a $4^\text{th}$ order exceptional point and all four eigenstates coalesce into one eigenstate of energy $E=0$.

Fig.~\ref{long}.b shows where the regions in the $r-c$ plane for which these types of exceptional points exist.

Note that when $b=0$ in Eq.~\ref{ab} and $a\neq 0$, the energies are $\pm \sqrt{2a}$ and $0$. The presence of such points in the Brillouin zone will not affect calculating the $2D$ Zak phase, as it can simply be defined using the states of energy $E=\pm \sqrt{a+\sqrt{a^2-b}}$, not the states of $E=\pm \sqrt{a-\sqrt{a^2-b}}$ since the latter can lead to coalescing of two energies at $E=0$ when $b=0$. Therefore, this case will not be of interest in this paper. 

\subsection{\texorpdfstring{$2^\text{nd}$}{2nd} Order Exceptional Points}

Consider the case when $2^\text{nd}$ order exceptional points exist, i.e. when $\Phi\equiv \Phi$ takes the value $0$ somewhere in the Brillouin zone. This is shown in the Fig.~\ref{long}.b as the combined red and blue regions.

In App.~\ref{app1}, it is shown that $\Phi$ has no degenerate critical points, except at $(\pi,\pi)$ when $r=c$. This means that $\Phi$ has no degenerate critical points when there exist exceptional points, except at the phase boundary $r=c=\frac{1}{2}$. When $\Phi$ has no nondegenerate critical points, it is a Morse function and, therefore, results from Morse theory can be used \cite{milnor}.

\begin{figure*}
\centering\includegraphics[width=\linewidth]{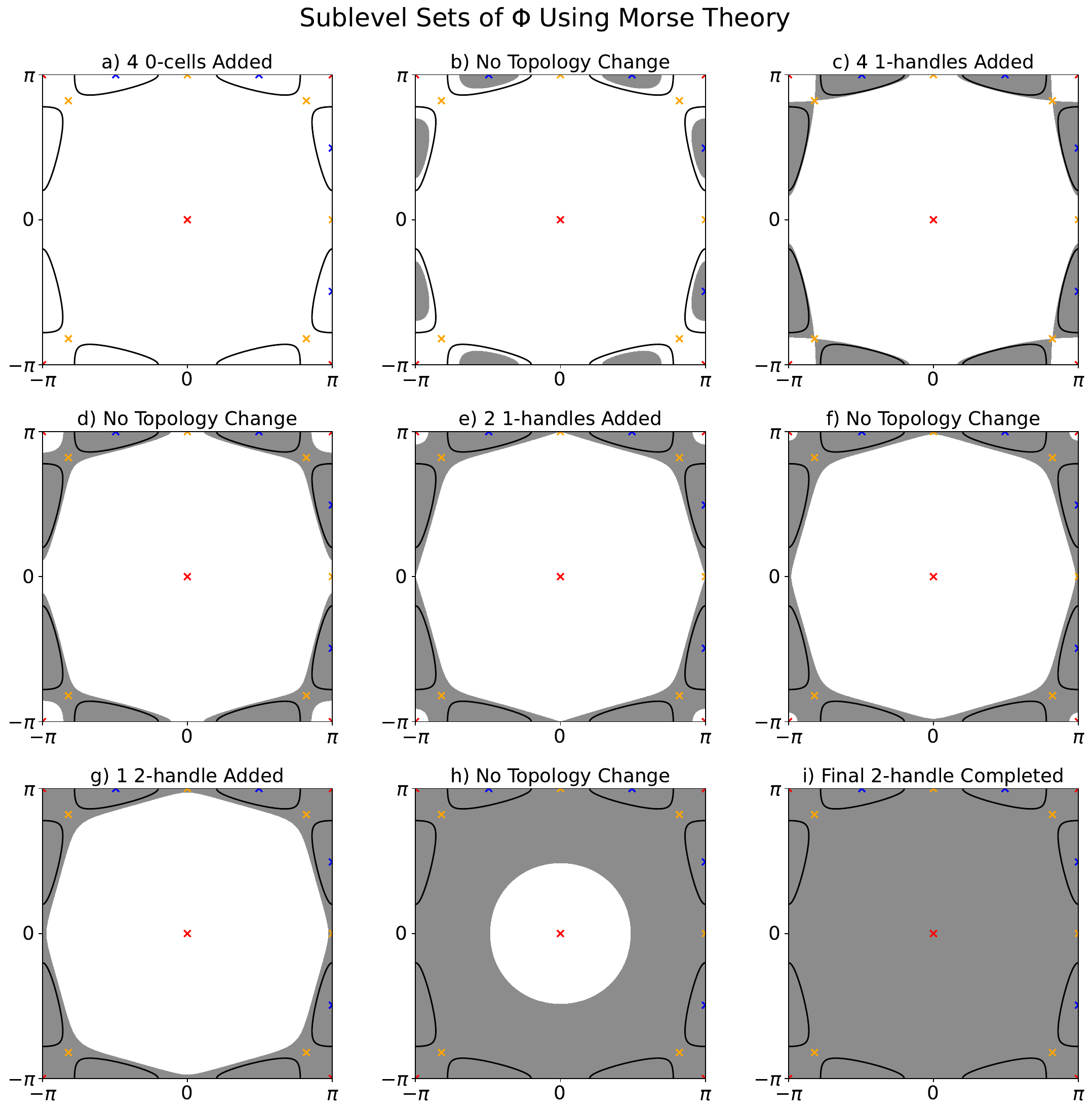}
\caption{An illustration of how the sublevel set $\Phi^{-1}(-\infty,c]$ changes as $c$ is increased. For every critical point of index $n$, an $n$-handle is added. i.e. a $2$-handle is added at each maximum (red points), a $1$-handle is added at each saddle (orange crosses) and a $0$-handle is added at each minimum (blue crosses).}
\label{morsefig}
\end{figure*}

Let the critical points of $\Phi$ be at $c_i$, where $c_i<c_{i+1}$. Since $\Phi$ is a Morse function, Morse theory implies that the sublevel set $\Phi^{-1}(-\infty,c]$ is attached to an $n$-handle at every critical point $c$ of index $n$, i.e. (number of negative eigenvalues of the Hessian). This result will be used to determine the different topologies of $\Phi^{-1}(-\infty,c]$ as $c$ is increased, with the different topologies shown in Fig.~\ref{morsefig}. Note that $\Phi^{-1}(0)$ is precisely where there are $2^\text{nd}$ order exceptional points.

In App.~\ref{app1}, it is shown that $\Phi$ has four minima where $\Phi=c_1$, four saddle points where $\Phi=c_2$, two saddles at $(0,\pi)$ and $(\pi,0)$ where $\Phi=c_3$, a maximum at $(\pi,\pi)$ where $\Phi=c_4$ and a maximum at $(0,0)$ where $\Phi=c_5$.
$\Phi^{-1}(-\infty,c_1+\epsilon]$ is four $0$-cells added to the empty set. There are no critical values between $c_1$ and $0$, so the topology of $\Phi^{-1}(-\infty,0]$, is unchanged and deformation retracts onto $\Phi^{-1}(c_1)$.

There are four saddle points for which $\Phi=c_2$. For small $\epsilon$, $\Phi^{-1}(-\infty,c_2-\epsilon] $ is still topologically four disks then $\Phi^{-1}(-\infty,c_2+\epsilon]$ is four disks with four $1$-handles attached.

There are then two $1$-handles attached due to the saddle points at $(0,\pi)$ and $(\pi,0)$. 

There is then one $2$-handle added due to the maximum at $(\pi,\pi)$.

Note that this discussion is in agreement with the Poincare-Hopf theorem applied to a Morse function on a torus, which implies that the number of saddle points of $\Phi$ must be equal to the number of non-saddle critical points of $\Phi$, both of which are $6$ here.

\subsection{\texorpdfstring{$4^\text{th}$}{4th} Order Exceptional Points}

\label{secep4}

% $b=0$ when 
% \begin{equation}
%     \cos(k_x)-\cos(k_y)= \pm \frac{(r^2-(1-r)^2)(c^2-(1-c)^2)}{2rc(1-r)(1-c)}.
% \end{equation}

% When 
% \begin{equation}
%     |(1-2r)(1-2c)| = 4rc(1-r)(1-c),
% \end{equation}
% $b=0$ at $(0,\pi)$ and $(\pi,0)$.

% When 
% \begin{equation}
%     |(1-2r)(1-2c)| \leq 4rc(1-r)(1-c),
% \end{equation}
% these the zero sets of $b$ split into two non-intersecting contractible loops within the Brillouin zone.

Consider the case when $4^\text{th}$ order exceptional points exist, which is the blue region in Fig.~\ref{long}.b. $4^\text{th}$ order exceptional points exist at eight points in the Brillouin zone $(\arccos(t_{\pm}), \arccos(t_{\mp}))$ for which all four eigenvalues of the Hamiltonian are zero where
\begin{equation}
    t_{\pm} \equiv \frac{-4c^2(1-c)^2\left[r^2+(1-r)^2\right] \pm (2r-1)\left[c^4-(1-c)^4\right]}{4r(1-r)c(1-c)\left[c^2+(1-c)^2\right]}.
\end{equation}
They exist for
\begin{equation}
\begin{split}
    &\left|r-\frac{1}{2}\right|\leq \left|w-\frac{1}{2}\right|,\\
    &w \equiv \frac{(4c^3-8c^2+6c-1)}{4c(1-c)}\\
   &+\frac{(1-2c)\sqrt{{8c^4-16c^3+12c^2-4c+1}}}{4c(1-c)}.
\end{split}
\end{equation}
When this bound is obeyed there are eight $4^\text{th}$ order exceptional points, which appear in pairs on each of the four loops of $2^\text{nd}$ order exceptional points. Note that when one of the bounds is attained precisely, the pairs of points on each loop meet each other on the boundary of the Brillouin zone (centered at $(0,0)$), resulting in four $4^\text{th}$ order exceptional points.

When $E=0$, either $|A|=|D|$ or $|B|=|C|$ holds. The second case is just a reflection of the first case in the line $k_x=k_y$, so without loss of generality, consider $|A|=|D|$. Although the phases are ill-defined at $E=0$, $|\mu|=|\mu''|=0$ and $|\mu'|=|\mu'''|\neq 0$, which gives well-defined $r'=r'''=0$ and $r=r''=\frac{1}{\sqrt{2}}$. Instead of using $\mu$ to determine $\theta$, one can determine $\theta$ directly from $e^{-2i\theta}=\tau=-\frac{D}{A^\ast}$. Therefore, $\theta$ has a well-defined limit at the exceptional point. However, it is important to note that $\theta'$ and $\theta'''$ are ill-defined.

 \begin{figure}
    \centering
    \includegraphics[width=\columnwidth]{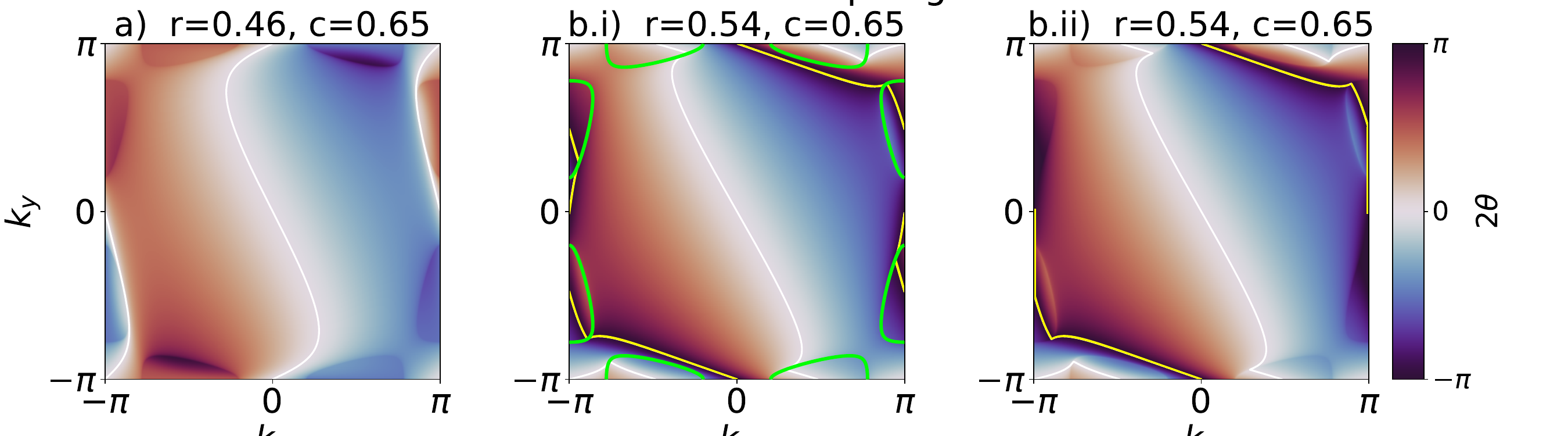}
    \caption{Plots of $2\theta(k_x,k_y)$ in the complex phase when loops of exceptional points exist. a) For $r<\frac{1}{2}$, there is still a globally smooth $2\theta(k_x,k_y)$ over the entire Brillouin zone with no discontinuities and it never reaches $\pm \pi$. This implies zero winding numbers and zero Zak phase. This is the same as the simpler case without any exceptional points. b) For $r>\frac{1}{2}$, there is a $2\pi$-discontinuitity in $2\theta$ that looks slightly different from the topological regime without exceptional points in Fig.~\ref{Fig41}. However, the discontinuity for each $k_x$ and $k_y$ slice in $2\theta(k_x,k_y)$ is still $2\pi$, leading to the same winding numbers. c) A shift by $2\pi$ in parts of the Brillouin zone (i.e. a gauge shift) shows that this is in fact exactly the same case as Fig.~\ref{Fig41} and the same arguments apply.}
    \label{Fig4}
\end{figure}

\subsection{Trivial Winding}

Within the contractible loops for which $\Phi \leq 0$, Eq.~\ref{taufrac} becomes
\begin{equation}
\begin{split}
    &\tau=e^{-2i\theta}=\frac{F+iG}{H-iJ}=\frac{H+iJ}{F-iG}\\
    \implies &2\theta=\text{phase}(-iH-J).
    \end{split}
\end{equation}
It is proven in App.~\ref{HJ} that, for $r<\frac{1}{2}$, there are no points in the Brillouin zone for which $H=J=0$. Therefore, within the regions for which $\Phi \leq 0$, $2\theta$ is a smooth map from a contractible space to $S^1$, which must therefore be null-homotopic and therefore have trivial winding number.

When $r<\frac{1}{2}$, outside the regions where $\Phi<0$, the same arguments as when there were no exceptional points hold. $2\theta$ does not reach $\pi$ anywhere in the Brillouin zone since there are no lines for which $G-J=0$ and $H<0$. The zero sets of $G-J=0$ do not intersect the regions where $\Phi \leq0$ except tangentially since $G-J=0$ implies $\Phi=H^2>0$. A plot of $2\theta$ with trivial winding number is given in Fig.~\ref{Fig4}.a.

\subsection{Nontrivial Winding}

Here, it is proven that, for $r>\frac{1}{2}$, the winding number of $\tau=e^{-2i\theta}$ is the same whether or not there are exceptional points. 

When there are exceptional points, there are four contractible regions within which $\Phi\leq 0$. There are two noncontractible loops along which $G-J=0$ for the same reasons as in the case without exceptional points (and no contractible loops, as proven in App.~\ref{g}). $\Phi=0$ tangentially touches $G-J=0$ where $H=0$ non-tangentially intersects $\Phi=0$. Note that it is proven in App.~\ref{H} that $H=0$ and $\nabla H=0$ never simultaneously hold in the Brillouin zone (except for a special case when the zero sets of $H$ are just formed of two isolated in the Brillouin zone), which implies that the zero sets of $H$ are a disjoint set of closed curves.

When the line along which $G-J=0$ and $H<0$ (so $\tau=-1$) tangentially intersects the contractible loop along which $\Phi=0$, $H=0$ at the intersection point and furthermore there is a line for which $H=0$ through the region within which $\Phi\leq0$ to another boundary point of the region. $\tau=-1$ through this entire line. At this second boundary point, $H=0$ intersects non-tangentially the line along which $\Phi=0$ whereas the line along which $G-J=0$ intersects the line along which $\Phi=0$ tangentially. Along one of the directions of the line for which $G-J=0$ from this point, $H<0$ and $\tau=-1$. This pattern then continues till the line $\tau=-1$ returns to the starting point. 

$H=0$ twice along the line $k_x=\pi$, each of which connects the left half of the Brillouin zone to the right half of the Brillouin zone (one in the top half and one in the bottom half). These ensure that the loops along which $\tau=-1$ are still noncontractible when they go through the regions where $\Phi \leq 0$. 

A similar argument gives a second noncontractible loop along which $\tau=+1$.

These arguments are quite involved and it is much easier to understand them graphically. In Fig.~\ref{Fig4}.b, the line along which $\tau=-1$ is yellow and the the lines along which $\Phi=0$ are green. Yellow lines are those for which either $G-J=0$ and $H<0$ outside the regions bounded by green or $H=0$ within the regions bounded by green.

The place where $2\theta$ jumps by $2\pi$ can be shifted arbitrarily and Fig.~\ref{Fig4}.b. shows how this case is indeed exactly the same as when there are no exceptional points.

\subsection{Generalized Invariant}

At exceptional points, the conventional band structure description breaks down. However, it was shown that the winding numbers of $2\theta$ for $r<\frac{1}{2}$ and $r>\frac{1}{2}$ are preserved away from the exceptional points. At a $4^\text{th}$ order exceptional point, either $\theta$ is well-defined and $\theta'$ is not or vice versa.

It is possible to define a topological invariant on the Brillouin zone with all exceptional points removed. This was considered for, for example, a type of Chern invariant on a punctured Brillouin zone in \cite{punctured}. This allows the usual Zak phase to be defined on this reduced Brillouin zone. 

It would also be expected that it could be dealt with by integrating around an infinitesimal deformation in complex momentum around the exceptional point, as discussed for other models in \cite{deformcontour}.

Another option would be to give the Hamiltonian a symmetry-preserving deformation which removes the exceptional points but preserves the 2D Zak phase.

\section{Edge Modes}

In the 2D Hermitian SSH model, the existence of a topologically protected phase with edge modes has been established \cite{wakabayashi2}. The purpose of this section is to give an analytical derivation of this in order to provide a basis from which to explore the non-Hermitian case more precisely.

Here, real momentum is considered since the Generalized Brillouin Zone was shown to be the same as the usual Brillouin zone in Sec.~\ref{secgbz}. Similarly, consider solutions on a semi-infinite lattice with finite $x$-direction of length $(N_x+1)$ of the form
\begin{equation}
\begin{pmatrix}
    \psi_{m,A}\\
    \psi_{m,B}\\
    \psi_{m,C}\\
    \psi_{m,D}
\end{pmatrix}
=
\begin{pmatrix}
    C_A(e^{ik_x m}-e^{-ik_x m})\\
    C_B(e^{ik_x m}-e^{-ik_x m})\\
    C_C(e^{ik_x m}-Z^2e^{-ik_x m})\\
    C_D(e^{ik_x m}-Z^2e^{-ik_x m})
\end{pmatrix}
e^{ik_y}
\end{equation}
where solutions are assumed to be superpositions of forwards and backwards travelling waves in the finite direction. Relative factors of $Z\equiv e^{ik_x(N_x+1)}$ arise from imposing periodicity in the finite direction.

It is shown in App.~\ref{edgu} that imposing the 1D equations of motion leads to the condition
\begin{equation}
    \begin{split}
        &16r^2c(1-c)\sin^2(k_x)\\
        &\left[c(1-r)\sin(k_x(N_x+1))+(1-c)r\sin(k_x(N_x+2))\right]\\
        &\left[(1-c)(1-r)\sin(k_x(N_x+1))+cr\sin(k_x(N_x+2))\right]\\
        &-2(1-2r)^2(1-2c)^2\cos(2k_x(N_x+1))=0.
    \end{split} \label{end} 
\end{equation}

In the Hermitian limit $c=\frac{1}{2}$, this is
\begin{equation}
    \sin(k_x)\left[(1-r)\sin(k_x(N_x+1))+r\sin(k_x(N_x+2))\right]=0.
\end{equation} 
Since this equation is symmetric under $k_x \rightarrow -k_x$, consider the domain $k_x\in [0,\pi]$. The first factor gives solutions at the endpoints of this domain. Note that $k_x=0,\pi,2\pi$ are unphysical solutions since the wavefunction becomes zero.

The second factor gives $(N_x)$ physical solutions in this domain for $r=1$ and $(N_x-1)$ solutions in this domain for $r=0$. The second factor has derivative $1$ at the zero at $k_x=0$ and derivative $(-1)^{N_x+1}\left[(N_x+1)-r(2N_x+3)\right]$. So the critical value of $r$ where the number of physical solutions changes is at 
\begin{equation}
    r'=\frac{N_x+1}{2N_x+3}.
\end{equation}
There is a full proof of this statement in App.~\ref{j}. This is in agreement with the numerical results in Ref.~\cite{wakabayashi2} and gives an analytical expression for the critical value of $r$.

Eq.~\ref{end} is made much more complex by the addition of the second term, which makes a similar analytical counting of solutions much more difficult. However, making choosing certain non-Hermitian deformations of the Hermitian limit might give a new insight into exactly how edge modes transition into the non-Hermitian regime.

\section{Conclusions and Outlook}

In this paper, the 2D Zak phase was explicitly calculated for a non-Hermitian SSH-type system. A natural extension of this work would be to consider models in higher dimensions and with generalizations such as longer-range transitions. It might also be interesting to revisit the in-depth analysis done of the 1D non-Hermitian SSH model \cite{lieu1} to see the similarities and differences to this 2D model.

Morse theory was used to prove that exceptional points appear only in contractible loops or at isolated points in momentum space. Morse-theoretic language could have been used in many other calculations in this paper, especially since analyzing the 2D Zak phase involved calculating a winding number for a 'phase' function containing Morse functions. As characterizing topological systems often relies on the topological structure of functions on a $d$-dimensional torus, Morse theory could prove to be useful in tackling a wider range of problems analytically than done previously \cite{morse1,morse2,morse3,morse4,michel1}. A systematic study of Morse theory applied to studying topological systems would be a useful avenue of future investigation.

An important open question is how to extend topological invariants to regimes where there are certain types of exceptional points. This would be interesting from a mathematical perspective in seeing what topology remains. The example given in this paper might also suggest ways to give a physical interpretation to topological structure that is exists with exceptional points.

An analytical solution for the Hermitian case of edge states was also shown. This was previously only shown numerically \cite{wakabayashi2}. This closed form solution offers a new insight into edge modes in Hermitian systems, as well as improves an understanding of the structure of edge modes in the non-Hermitian regime.

The 2D Zak phase is a weak topological invariant in the sense that it is defined on lower-dimensional submanifolds of the Brillouin zone, which can be well-understood physically in the Hermitian case \cite{weak1,weak2}, whereas a full framework to understand non-Hermitian weak invariants would be interesting to pursue.

% \begin{figure}
% \vspace{-2cm}
% \hspace{-2cm}
% \includegraphics{edg.png}
% \caption{Number of edge modes for N=40 system for varying (r,c).}
% \label{edg}
% \end{figure}
\newpage
\appendix
\section{Structure of \texorpdfstring{$\Phi$}{(a²-b)}}
\label{app1}
\begin{equation}
\begin{split}
    \Phi \equiv& \left(2c(1-c)[r^2+(1-r)^2]\right.\\
   &\left.+r(1-r)[c^2+(1-c)^2][\cos(k_x)+\cos(k_y)]\right)^2 \\
    &-4r^2(1-r)^2 c^2 (1-c)^2 [\cos(k_x)-\cos(k_y)]^2 \\
    &+ [r^2-(1-r)^2]^2[c^2-(1-c)^2]^2 \label{A1}
    \end{split}
\end{equation}

Setting the derivative of Eq.~\ref{A1} with respect to $k_x$ equal to zero gives
\begin{equation}
\begin{split}
   & -\sin(k_x)\left[2r^2(1-r)^2[c^2-(1-c)^2]^2\cos(k_x)\right.\\
   &\left.+2r^2(1-r)^2[[c^2+(1-c)^2]^2+  
     4c^2(1-c)^2] \cos(k_y) \right.\\
     &\left.+4r(1-r)c(1-c)[r^2+(1-r)^2][c^2+(1-c)^2] \right]=0 .\label{A2a}
\end{split}
\end{equation}
Similarly, setting the derivative with respect to $k_y$ equal to zero gives the same equation as Eq.~\ref{A2a} with $k_x$ and $k_y$ interchanged.
Consider the critical points where both derivatives are zero, where for both expressions, at least one out of two multiplicative factors must be zero. This happens when $\sin(k_x)\neq0$ and $\sin(k_y)\neq$ only if
\begin{equation}
    \cos(k_x)=\cos(k_y)=-\frac{c(1-c)[r^2+(1-r)^2]}{r(1-r)[c^2+(1-c)^2]}.
\end{equation}
Substituting this into Eq.~\ref{A1} gives $\Phi=[r^2-(1-r)^2]^2[c^2-(1-c)^2]^2 > 0$ with an indefinite Hessian, so it is a positive saddle point.

Secondly, assume that $\sin(k_x)=0$ and $\sin(k_y)=0$. $(0,0)$ is a positive maximum. $(\pi,0)$ is a positive saddle point. $(\pi,\pi)$ is either a positive minimum or maximum, except when $r=\pm c$.

When $\sin(k_x)=0$ and $\sin(k_y) \neq 0$. $k_x=0$ would imply $\cos(k_y)<-1$ but $k_x=\pi$ implies $k_y=\pm u$ where
\begin{equation}
    \cos(u)\equiv\left[1-\frac{2c(1-c)[2c^2-2c+2r^2-2r+1]}{r(1-r)[c^2-(1-c)^2]^2}\right]. \label{plus}
\end{equation}

The Hessian is positive definite here, implying there exist minima at $(\pm \pi, u)$ and $(u,\pm \pi)$. These are the only points at which it is possible to have a negative minimum.
\section{Symmetries}
\label{symmetriesapp}

\subsection{\texorpdfstring{$C_4$}{C4} symmetry}

The $C_4$ symmetry for the gauge choice in  Eq.~\ref{topgauge}
\begin{equation}
    U_{C_4}|R(k_x,k_y)\rangle = e^{i(\theta'''-\theta'')}|R(-k_y,k_x)\rangle,
\end{equation}
which implies that
\begin{equation}
    \begin{split}
&Z_{k_x}=\frac{1}{2\pi}\int_{\text{B.Z.}}\langle R(k_x,k_y)|-i\partial_{k_x}|R(k_x,k_y)\rangle\\
=&\frac{1}{2\pi}\int_{\text{B.Z.}}\langle R(-k_y,k_x)|(U_{C_4})(-i\partial_{k_x})(U_{C_4}^{-1})|R(-k_y,k_x)\rangle\\
&+\frac{1}{2\pi}\int_{B.Z.}\langle R(-k_y,k_x)|R(-k_y,k_x\rangle \partial_{k_x}\left(\theta'''-\theta''\right)\\
=&\frac{1}{2\pi}\int_{\text{B.Z.}}\langle R(-k_y,k_x)|(-i\partial_{k_x})|R(-k_y,k_x)\rangle\\
&+\frac{1}{2\pi}\int_{B.Z.}\partial_{k_x}\left(\theta'''-\theta''\right)\\
=&\frac{1}{2\pi}\int_{\text{B.Z.}}\langle R(k_x,k_y)|(-i\partial_{k_y})|R(k_x,k_y)\rangle\\
&+\frac{1}{2\pi}\int_{B.Z.}\partial_{k_x}\left(\theta'''-\theta''\right)\\
=&Z_{k_y}+\frac{1}{2\pi}\int_{B.Z.}\partial_{k_x}\left(\theta-\theta'\right)
\end{split}
\end{equation}

So 
\begin{equation}
\begin{split}
    Z_{k_y}=-Z_{k_x}+(n_y-n_x)\pi \mod 2\pi.
\end{split}
\end{equation}

This implies that $Z_{k_i}$ can only take the values $0$ or $\pi$ (mod $2\pi$). For this particular model, $n_x=n_y$, so the $C_4$ symmetry implies that $Z_{k_x}=Z_{k_y} \mod 2\pi$.
\subsection{Generalized Pseudo-Hermitian Symmetry}
Since
\begin{equation}
    P|R(k_x,k_y)\rangle = e^{i\left(\theta(-k_y,-k_x)-\theta(-k_x,-k_y))\right)}|L(k_y,k_x)\rangle,
\end{equation}
implies that (assuming, as in the case of this model, that $n_x=n_y$)
\begin{equation}
    \begin{split}
 Z_{k_x} = &\frac{1}{2\pi}\int_{\text{B.Z.}}\langle R(k_x,k_y)|-i\partial_{k_x}|R(k_x,k_y)\rangle\\
=&\frac{1}{2\pi}\int_{\text{B.Z.}}\langle L(k_y,k_x)|(P)(-i\partial_{k_x})(P^{-1})|L(k_y,k_x)\rangle\\
&+\frac{1}{2\pi}\int_{B.Z.}\partial_{k_x}\left(\theta(-k_y,-k_x)-\theta(-k_x,-k_y)\right)\\
=&\frac{1}{2\pi}\int_{\text{B.Z.}}\langle L(k_y,k_x)|-i\partial_{k_x}|L(k_y,k_x)\rangle\\
&+(n_x-n_y)\pi\\
=&\frac{1}{2\pi}\int_{\text{B.Z.}}\langle L(k_x,k_y)|-i\partial_{k_y}|L(k_x,k_y)\rangle\\
 \equiv &Z^{LL}_{k_y}.
    \end{split}
\end{equation}

So
\begin{equation}
\begin{split}
    &Z_{k_x}=Z^{LL}_{k_y}+(n_x-n_y)\pi,\\
    &Z_{k_y}=Z^{LL}_{k_x}+(n_y-n_x)\pi.
\end{split}
\end{equation}

For this model, since $n_x=n_y$, $Z_{k_x}=Z_{k_y}^{LL}$ and $Z_{k_y}=Z_{k_x}^{LL}$. Using the conclusion derived from $C_4$ symmetry (and $n_x=n_y$) that $Z_{k_x}=Z_{k_y}$ leads to the conclusion that $Z_{k_i}=Z_{k_i}^{LL}$, i.e. it does not matter whether left or right eigenstates are used to define the Zak phase.

\section{Choice of Energy Band}
\label{appband}
Symmetries can be used to link the 2D Zak phase calculated with eigenstates of different energy bands.
Sublattice symmetry means that
\begin{equation}
    |R_a(k_x,k_y)\rangle=\mathcal{S}_+|R_b(k_x,k_y)\rangle,
\end{equation}
for corresponding eigenvalues
\begin{equation}
    E_a=-E_b
\end{equation}
so
\begin{equation}
    \begin{split}
&\frac{1}{2\pi}\int_{\text{B.Z.}}\langle R_a(k_x,k_y)|\mathcal{S}_+(-i\partial_{k_x})(\mathcal{S}_+)^{-1}|R_a(k_x,k_y)\rangle\\
=&\frac{1}{2\pi}\int_{\text{B.Z.}}\langle R_b(k_x,k_y)| -i\partial_{k_x}|R_b(k_x,k_y)\rangle \\
=&\frac{1}{2\pi}\int_{\text{B.Z.}}\langle R_b(k_x,k_y)|-i\partial_{k_x}|R_b(k_x,k_y)\rangle,  
\end{split}
\end{equation}
so Zak phases are unchanged when a state of energy $E$ is swapped with a state of energy $-E$.
Time-reversal symmetry means that
\begin{equation}
    |R_a(k_x,k_y)\rangle=|R_b(-k_x,-k_y)\rangle^\ast,
\end{equation}
for corresponding eigenvalues
\begin{equation}
    E_a=E_b^\ast
\end{equation}
so
\begin{equation}
    \begin{split}
&\frac{1}{2\pi}\int_{\text{B.Z.}}\langle R_a(k_x,k_y)|\mathcal{T}^+(-i\partial_{k_x})(\mathcal{T}^+)^{-1}|R_a(k_x,k_y)\rangle\\
=&\frac{1}{2\pi}\int_{\text{B.Z.}}\langle R_b(-k_x,-k_y)|^\ast (-i\partial_{k_x})|R_b(-k_x,-k_y)\rangle^\ast \\
=&\frac{1}{2\pi}\int_{\text{B.Z.}}\langle R_b(k_x,k_y)|^\ast (+i\partial_{k_x})|R_b(k_x,k_y)\rangle^\ast,  
\end{split}
\end{equation}
so Zak phases are unchanged when a state of energy $E$ is swapped with a state of energy $E^\ast$.

Since energies come in complex conjugate pairs and pairs of opposite sign, the 2D Zak phase is identical when calculated with any one of the four states. This is not generally true for non-Hermitian systems.

% \section{Proof that \texorpdfstring{$|\tau|$}{|tau|} has no poles}
% \label{apptau}
% When $\Phi<0$, we get
% \begin{equation}
%     |\tau|^2=\frac{\mp \sqrt{\Phi}-i(\gamma_\text{in}\gamma_\text{ex}-\gamma'_\text{in}\gamma'_\text{ex})(\sin(k_x) - \sin(k_y))} {\pm \sqrt{\Phi}-i(\gamma_\text{in}\gamma_\text{ex}-\gamma'_\text{in}\gamma'_\text{ex})(\sin(k_x) - \sin(k_y))}
% \end{equation}

% We can analytically see whether $|\tau|$ has zeros or poles by finding the zeros of the numerator and the denominator, i.e. the solutions of
% \begin{equation}
%     \Phi+r^2(1-r)^2\left[c^2-(1-c)^2\right]^2(\sin(k_x)- \sin(k_y))^2=0
% \end{equation}
% This function has minima at two points in the Brillouin zone: $ \pm (v,  v)$ where
% \begin{equation}
%     \cos(v)= \frac{-c(1-c)(r^2+(1-r)^2}{r(1-r)(c^2+(1-c)^2)}.
% \end{equation}
% The function takes the value there at
% \begin{equation}
% \begin{split}
%     &\Phi+r^2(1-r)^2\left[c^2-(1-c)^2\right]^2(\sin(k_x) - \sin(k_y))^2\\
%     =&(2r-1)^2(2c-1)^2 \geq 0.
%     \end{split}
% \end{equation}
% with this quantity never reaching zero except $(r,c)=(\frac{1}{2},\frac{1}{2})$.

% Note that $\tau$ can never have zero numerator and denominator simultaneously since $F=G=H=J=0$ has no solutions. 

\section{\texorpdfstring{Topology of Zero Sets of $G-J=0$}{g-=0}}
\label{g}
$G-J$ is
\begin{equation}
\begin{split}
    G-J \equiv& -\sin(k_x+k_y)r^2(c^2+(1-c)^2)\\
    &+\sin(k_x)r(1-r)(2c^2-1)\\
    &+\sin(k_y)r(1-r)(2c^2-4c+1) .
    \end{split}
\end{equation}

If $\partial_{k_x}(G-J)=\partial_{k_y}(G-J)$,
\begin{equation}
\begin{split}
    &(1-r)(2c^2-1)\cos(k_x)=(1-r)(2c^2-4c+1)\cos (k_y)\\
    &=r(c^2+(1-c)^2)\cos(k_x+k_y).
    \end{split}
\end{equation}

If also $G-J=0$, away from special points when $r,c$ take values $(0,\frac{1}{2},1)$, the only solution is when
\begin{equation}
\begin{split}
    &k_x=\arccos\left(\frac{(1-r)(2c^2-4c+1)}{r(c^2+(1-c)^2)}\right),\\
    &k_y=\arccos \left( \frac{(1-r)(2c^2-1)}{r(c^2+(1-c)^2)}\right). \label{minusone}
    \end{split}
\end{equation}

$G-J$ is zero with zero gradient if and only if Eqs.~\ref{minusone} hold and one coordinate is $\pi$ or $-\pi$. This only happens for $r=\frac{2c^2-4c+1}{-2c}$ and $c>\frac{1}{2}$ or for $r=\frac{2c^2-1}{-2(1-c)}$ and $c<\frac{1}{2}$. Apart from this special case, which is described in App.~\ref{spec}, $G-J$ is never zero with zero gradient. 

Assume there is a contractible loop on which $G-J=0$ within the Brillouin zone and the special case just described does not hold. Then Eqs.~\ref{minusone} are points at which the gradient of that loop are $-1$, each of which lies in a quadrant of the Brillouin zone. $G-J \neq 0$ on the lines $k_x=0$, $k_x=\pi$ and $k_y=\pi$ except at $(0,0)$, $(0,\pi)$, $(\pi,0)$ and $(\pi,\pi)$. Therefore, any contractible loop of zeros of $G-J=0$ must lie completely within one quadrant of the Brillouin zone, excluding the boundary. Since there is only one point at which the loop reaches gradient $-1$, this is a contradiction. Therefore, there are never any contractible loops for which $G-J=0$.

\section{Special Case}
\label{spec}
Consider $r=\frac{2c^2-4c+1}{-2c}$ and $c>\frac{1}{2}$ or when $r=\frac{2c^2-1}{-2(1-c)}$ and $c<\frac{1}{2}$. Without loss of generality, consider the first case. Then $k_x=\pi$ implies $G-J=0$. 

\begin{equation}
\begin{split}
    \left(\pi,\arccos \left( \frac{(1-2c^2)}{(2c^2-4c+1)}\right)\right),
    \end{split}
\end{equation}
which always exists for $c>\frac{1}{2}$.

The determinant of the Hessian of $G-J$ is negative here so this is simply a saddle point where two lines cross.

\section{When \texorpdfstring{$H$}{H} is zero with zero gradient}
\label{H}
\begin{equation}
\begin{split}
    &H(k_x,k_y)\equiv(1-r)^2[c^2+(1-c)^2]\\
    &+2rc(1-r)(1-c)[\cos(k_x)+\cos(k_y)]\\
    &+r^2[c^2+(1-c)^2]\cos(k_x+k_y).
\end{split}
\end{equation}
 Similarly to the analysis for $G-J$, $H$ is smooth and is never zero with zero gradient, so level sets form closed loops in the Brillouin zone. Setting derivatives of this expression with respect to each variable to zero gives the condition
\begin{equation}
    \begin{split}
        H_{k_x}=&-2rc(1-r)(1-c)\sin(k_x)\\
        &-r^2[c^2+(1-c)^2]\sin(k_x+k_y)=0, \\
        H_{k_y}=&-2rc(1-r)(1-c)\sin(k_y)\\
        &-r^2[c^2+(1-c)^2]\sin(k_x+k_y)=0.
    \end{split}
\end{equation}
These conditions imply $\sin(k_x)=\sin(k_y)$. Substituting this back into each condition gives
\begin{equation}
\begin{split}
        &\sin(k_x)\left[2rc(1-r)(1-c)+r^2[c^2+(1-c)^2]\right.\\ 
        &\left.\left[\cos(k_x) \pm \cos(k_y)\right]\right]=0, 
        \end{split}
\end{equation}
and the same with $k_x$ and $k_y$ switched.
These conditions are solved with $\sin(k_x)=\sin(k_y)=0$ or
\begin{equation}
    k_x=k_y=\arccos\left(-\frac{(1-r)c(1-c)}{r[c^2+(1-c)^2]}\right).
\end{equation}
$H=0$ at this latter critical point only if $r=\frac{c(1-c)}{1-c(1-c)}$.

For $\sin(k_x)=\sin(k_y)=0$, $H(0,0)$ and $H(\pi,\pi)$ are both positive. $H(0,\pi)=H(\pi,0)=(1-2r)(c^2+(1-c)^2)$, which is nonzero for $r\neq \frac{1}{2}$.

Therefore, in the Brillouin zone, $H$ is zero with zero gradient only when $r=\frac{c(1-c)}{1-c(1-c)}$. When this identity holds, there are two points in the Brillouin zone for which $H=0$.

\section{Proof that \texorpdfstring{$H=J=0$}{H=J=0} nowhere in the Brillouin zone for \texorpdfstring{$r<\frac{1}{2}$}{r<1/2}}
\label{HJ}

For $r<\frac{1}{2}$, $H(k_x,0) \neq 0$ so there are no noncontractible loops of zeros of $H$. The gradient of the zero contour of $H$ reaches $-1$ only along the line $k_x=k_y$. 

When $r<\frac{\sqrt{c^4+(1-c)^4}}{c^2+(1-c)^2+\sqrt{c^4+(1-c)^4}}$, $H$ is nonzero in the whole Brillouin zone. 

When $r=\frac{\sqrt{c^4+(1-c)^4}}{c^2+(1-c)^2+\sqrt{c^4+(1-c)^4}}$, there are two points in the Brillouin zone where $H=0$. $J>0$ at both these points, so $H=J=0$ nowehere in the Brillouin zone.

When $\frac{\sqrt{c^4+(1-c)^4}}{c^2+(1-c)^2+\sqrt{c^4+(1-c)^4}}<r<\frac{1}{2}$, there are four solutions to $H(k_x,k_x)=0$ at which 
\begin{equation}
\begin{split}
    &\cos(k_x)=\frac{-2c(1-r)(1-c)}{2r[c^2+(1-c)^2]}\\
    &\pm \frac{ \sqrt{2r^2[c^2+(1-c)^2]^2-2(1-r)^2[c^4+(1-c)^4]}}{2r[c^2+(1-c)^2]}.
\end{split}
\end{equation}
Therefore, there is one contractible loop of $H=0$ in each of the upper right and lower left quadrant of the Brillouin zone. These four points where $H=0$ has gradient $-1$ lie above $k_y=-k_x+\frac{1}{2}v$, for $\cos(v)\equiv -\frac{2r(1-r)c(1-c)}{r^2[c^2+(1-c)^2]}$, and therefore the whole of the contractible loop lies above this line.

$J=0$ along the noncontractible loop $k_x=-k_y$. $J(k_x,0)=0$ only on this line and at $(0,\pi)$, which is part of a second noncontractible loop of zeros of $J$.

When $r<2c(1-c)$, the second noncontractible loop connects $(0,\pi)$ to $(\pi,0)$ twice in the upper left and bottom right quadrant, with $J$ and its gradient both zero only at $(u,-u)$ where $\cos(u)=-\frac{r^2[c^2+(1-c)^2]}{2r(1-r)c(1-c)}$. At $r=2c(1-c)$, $u=\pi$, and the second noncontractible loop is along $k_x=\pi$ and $k_y=\pi$.

When $r>2c(1-c)$, the second noncontractible loop does go through the upper right and bottom left quadrants, reaching gradient $1$ only at $\pm(v,v)$. Therefore, the zero contour of $J$ never goes above the line $k_y=-k_x+\frac{1}{2}v$. Therefore, the contours of $H=0$ never intersect the contours of $J=0$.

When $r=2c(1-c)$, $J=0$ along $k_x=\pi$, $k_y=\pi$, and $k_x=-k_y$, along which there are no points for which $H=0$.

\section{Edge mode equations}
\label{edgu}
The 1D equations of motion lead to the following conditions:
\begin{equation}
\begin{split}
    &0=E^4[-2Z^2-2]\\
    &+E^2\left[Z^2(2BD^\ast +2B^\ast D+2AC+AC^\ast +A^\ast C)\right.\\
    &\left. +(2BD^\ast +2B^\ast D+2A^\ast C^\ast +AC^\ast +A^\ast C)\right]\\
    &+\left[Z^2((|A|^2-|D|^2)(|B|^2-C^2)+(|C|^2-|B|^2)(|D|^2-A^2)\right.\\
    &\left. +(|A|^2-|D|^2)(|B|^2-{C^\ast }^2)+(|C|^2-|B|^2)(|D|^2-{A^\ast }^2) \right],
    \end{split}
\end{equation}
and
\begin{equation}
    \begin{split}
        &0=E^4(Z^4+4Z^2+1)-E^2\left[Z^4(BD^\ast +B^\ast D+2AC)\right.\\ 
        &\left. +Z^2(2(A+A^\ast )(C+C^\ast )+4BD^\ast +4B^\ast D)\right.\\
        &\left.+(2A^\ast C^\ast +BD^\ast +B^\ast D)\right]+Z^4(|D|^2-A^2)(|B|^2-C^2)\\
        &+Z^2(4|B|^2|D|^2-|B|^2(A+A^\ast )^2-|D|^2(C+C^\ast )^2\\
        &+(AC^\ast +A^\ast C)^2)+(|D|^2-{A^\ast }^2)(|B|^2-{C^\ast }^2)
    \end{split}
\end{equation}

The former of these is equivalent to 
\begin{equation}
    \begin{split}
       &0= Z^2\left[E^2(2AC-A^\ast C-AC^\ast )\right. \\
       &+\left.(|A|^2-|D|^2)(|B|^2-C^2)+(|C|^2-|B|^2)(|D|^2-A^2)\right.\\
       &\left.+2(|A|^2-|B|^2)(|C|^2-|D|^2)\right]\\
       &+\left[E^2(2A^\ast C^\ast -A^\ast C-AC^\ast )\right.\\
        &+(|A|^2-|D|^2)(|B|^2-{C^\ast }^2)+(|C|^2-|B|^2)(|D|^2-{A^\ast }^2) \\
        &\left.+2(|A|^2-|B|^2)(|C|^2-|D|^2)\right]. \label{u1}
    \end{split}
\end{equation}

The latter is equivalent to 
\begin{equation}
    \begin{split}
        &0=Z^4\left[E^2(A^\ast C+AC^\ast -2AC)\right.\\
        &\left.+(|D|^2-A^2)(|B|^2-C^2)-(|A|^2-|B|^2)(|C|^2-|D|^2)\right  ]\\
        &+Z^2\left[E^2(2A^\ast C+2AC^\ast -2AC-2A^\ast C^\ast )+4|B|^2|D|^2\right.\\
        &-|B|^2(A+A^\ast )^2-|D|^2(C+C^\ast )^2+(AC^\ast +A^\ast C)^2\\
        &\left.-4(|A|^2-|B|^2)(|C|^2-|D|^2)\right]\\
        &+\left[E^2(AC^\ast +A^\ast C-2A^\ast C^\ast )\right.\\
        &\left.+(|D|^2-{A^\ast }^2)(|B|^2-{C^\ast }^2)-(|A|^2-|B|^2)(|C|^2-|D|^2)\right] 
    \end{split} \label{u2}
\end{equation}

Adding Eq.\ref{u1} multiplied by $(Z^2+1)$ into Eq.\ref{u2} gives 
\begin{equation}
\begin{split}
    &Z^4\left[((|A|^2-A^2)(|C|^2-C^2)-(|A|^2-|C|^2)^2\right]\\
    &+Z^2\left[((|A|^2-{A^\ast }^2)(|C|^2-C^2)+((|A|^2-A^2)(|C|^2-{C^\ast }^2)\right]\\
    &+\left[((|A|^2-{A^\ast }^2)(|C|^2-{C^\ast }^2)-(|A|^2-|C|^2)^2\right]=0
    \end{split}
    \label{edge2}
\end{equation}

This is then equivalent to Eq.~\ref{end}.
%\begin{equation}
%\begin{split}
%    |A|^2-A^2&=(1-c)^2r^2(1-e^{2ik_x})+rc(1-r)(1-c)(e^{-ik_x}-e^{ik_x})\\
%    &=-2i \sin(k_x) (1-c)r\left[(1-r)c+(1-c)re^{ik_x}\right]\\
 %   |C|^2-C^2 &= -2i \sin(k_x) cr\left[(1-r)(1-c)+cre^{ik_x}\right]\\
  %  |C|^2-C^{*2} &= 2i \sin(k_x) cr\left[(1-r)(1-c)+cre^{-%ik_x}\right]\\
  %  &4\sin^2(k_x)rc(1-r)(1-c)
  %  \end{split}
%\end{equation}
\section{Critical value of \texorpdfstring{$r$}{r} in the Hermitian limit}
\label{j}
\begin{equation}
    j\equiv (1-r)\sin(k_x(N_x+1))+r\sin(k_x(N_x+2)).
\end{equation}

This is known as a real trigonometric polynomial of degree $N_x+2$. Such functions have at most $2(N_x+2)$ roots in the interval $[0,2\pi)$ unless it is the zero function. 

Consider $\frac{n \pi}{N_x+2} < \frac{n \pi}{N_x+1} < k_x < \frac{ (n+1) \pi}{N_x +2} < \frac{ (n+1) \pi}{N_x+1}$ for integers $0 < n < (N_x+1)$. Both terms in $j$ are the same sign and nonzero so $j$ has no zeros within these intervals. Note that the same argument holds for $0 < k_x<\frac{\pi}{N_x+2}$.

Consider $   \frac{ (n-1) \pi}{N_x+1}  <\frac{ n \pi}{N_x +2} < k_x < \frac{ n \pi}{N_x+1}<\frac{ (n+1) \pi}{N_x +2}$ for integer $0<n<(N_x+1)$. The sign of $j$ is opposite at $\frac{ n \pi}{N_x +2}$ and $\frac{ n \pi}{N_x+1}$, so by the intermediate value theorem the number of zeros in this interval is $1 \enspace \text{mod} \enspace 2$ (therefore at least one). This gives at least $N_x$ zeros in the interval $0<k_x<\frac{N_x \pi}{N_x+1}$. By symmetry, there are also at least $N_x$ zeros in the interval $\frac{(N_x+2) \pi}{N_x+1}<k_x<2 \pi$. Furthermore, the number of zeros from these intervals is collectively a multiple of $4$. There is also a zero at $0$.

Consider $\frac{N_x \pi}{N_x+1} < \pi < \frac{(N+2)\pi}{N_x+1}$. $j$ has the same sign at the endpoints of this interval. For $r=\frac{N_x+1}{2N_x+3}$, there is a double-zero at $k_x=\pi$. For $r>\frac{N_x+1}{2N_x+3}$, the derivative of $j$ at $\pi$ is the same sign as $j$ at the left endpoint of the interval, meaning there are at least $3$ zeros in this interval. It also means that the number of zeros is $3 \enspace \text{mod} \enspace 4$ to match the signs of $j$ and its derivative, whilst still maintaining $k_x \rightarrow -k_x$ symmetry.

It has therefore been established that for $r>\frac{N_x+1}{2N_x+3}$, there is the maximal number of roots $2N_x+4$. 

For $r=\frac{N_x+1}{2N_x+3}$, there are precisely $2N_x+3$ roots (there cannot be one more root since, by symmetry, it would have to be at $0$ or $\pi$ which is not possible).

For $r<\frac{N_x+1}{2N_x+3}$, $2N_x+2$ roots have been identified. Since it has been argued that the number of roots must be preserved $\text{mod} \enspace 4$, this is precisely the number of roots.

Removing the unphysical roots at $k_x=0,\pi,2\pi$ gives the required result.

\bibliography{paper2}% Produces the bibliography via BibTeX.

\end{document}